\documentclass[conference]{IEEEtran}
\IEEEoverridecommandlockouts
\usepackage{cite}
\usepackage{amsmath,amssymb,amsfonts}
\usepackage{algorithm2e}
\usepackage{setspace}
\usepackage{graphicx}
\usepackage{textcomp}
\usepackage{xcolor}
\usepackage{diagbox}
\usepackage{multirow}
\usepackage{subfigure}

\def\BibTeX{{\rm B\kern-.05em{\sc i\kern-.025em b}\kern-.08em
T\kern-.1667em\lower.7ex\hbox{E}\kern-.125emX}}
\begin{document}

\title{XBlock-ETH: Extracting and Exploring Blockchain Data From Ethereum}

\author{
\IEEEauthorblockN{Peilin Zheng}
\IEEEauthorblockA{\textit{School of Data and Computer Science} \\
\textit{Sun Yat-sen University}\\
Guangzhou, China\\
zhengpl3@mail2.sysu.edu.cn}
\and
\IEEEauthorblockN{Zibin Zheng$^*$}
\IEEEauthorblockA{\textit{School of Data and Computer Science} \\
\textit{Sun Yat-sen University}\\
Guangzhou, China\\
zhzibin@mail.sysu.edu.cn}
\and
\IEEEauthorblockN{Hong-Ning Dai}
\IEEEauthorblockA{\textit{Faculty of Information Technology
} \\
\textit{Macau University of Science and Technology}\\
Macau, SAR\\
hndai@ieee.org}
}

\maketitle

\begin{abstract}
Blockchain-based cryptocurrencies have received extensive attention recently. Massive data has been stored on permission-less blockchains. The analysis on massive blockchain data can bring huge business values. However, the lack of well-processed up-to-date blockchain datasets impedes big data analytics of blockchain data. To fill this gap, we collect and process the up-to-date on-chain data from Ethereum, which is one of the most popular permission-less blockchains. We name these well-processed Ethereum datasets as XBlock-ETH, which consists of the data of blockchain transactions, smart contracts, and cryptocurrencies (i.e., tokens). The basic statistics and exploration of these datasets are presented. We also outline the possible research opportunities. The datasets with the raw data and codes have been publicly released online. 


\end{abstract}

\section{Introduction}

Blockchain has attracted extensive attention from both academia and industry in the recent years. Among the diverse blockchain systems, substantial efforts have been made on the permission-less blockchain (or public blockchain) due to its decentralization \cite{zheng2016blockchain}. The idea of permission-less blockchain was firstly proposed and implemented on Bitcoin \cite{nakamoto2008bitcoin}. In a blockchain system, each peer holds a ledger being considered as a public tally that is essentially temper-resistant. Ethereum \cite{buterin2013ethereum} is another most popular permission-less blockchain system that enables Turing-complete smart contracts. 

The proliferation of blockchain systems has lead to the generation of massive amount of blockchain data. Take Bitcoin as an example. There are nearly 242 GB Bitcoin data by the third quarter of 2019 as reported by Statista (https://www.statista.com/). In this paper, we focus on the data of Ethereum rather than Bitcoin, since Ethereum provides richer data. For another example, more than 16,000,000 smart contracts are deployed on Ethereum. As the Ethereum community has published two token protocol to enable easier Initial Coin Offerings (so-called ICO) for users \cite{buterin2015erc20}, over 100,000 kinds of ERC20 token and 1,600 kinds of ERC721 token are available to be transferred on Ethereum where ERC stands for Ethereum Request for Comments. 

The massive blockchain data provides researchers with both huge business values and great opportunities \cite{hndai:blockchain-iot2019} due to openness, decentralization and temper-resistance of blockchain systems. Take business trading data as an example. In the past, it is difficult for researchers to obtain the real business trading data because of the privacy or ownership concerns of data owners. However, all the data in incumbent blockchain systems are all publicly available. Meanwhile, the blockchain data in permission-less blockchains can be accessed almost everywhere due to the decentralization of blockchain systems. Moreover, distributed consensus of blockchains also guarantees the temper-resistance of blockchain data. In addition to blockchain transactions, Ethereum (or its alternatives) also consists of both smart contracts and cryptocurrencies. Big data analytics of blockchain data can advance the developments in fraud detection of transactions, vulnerability detection of smart contracts and software development of smart contracts, etc. 

However, there are a number of challenges in big data analytics of blockchain data, especially in Ethereum: \textbf{(1) Difficulty in data synchronization at Blockchain peer.} Due to the bulky size of blockchain, it takes a long period to fully synchronize entire blockchain data at a node (i.e., a peer) newly connected with the blockchain. For example, it takes more than one week and over 500 GB storage space to fully synchronize the entire Ethereum at a peer. The high expenditure of massive storage space and network bandwidth due to blockchain data synchronization impedes the analysis of blockchain data. \textbf{(2) Challenge in blockchain data extraction and procession.} Blockchain data is stored at clients in heterogeneous and complex data structures, which cannot be directly analyzed. Meanwhile, the underlying blockchain data is either binary or encrypted. Thus, it is a necessity to extract and process binary and encrypted blockchain data so as to obtain valuable information. However, it is non-trivial to process heterogeneous blockchain data since conventional data analytic methods may not work for this type of data.
\textbf{(3) Absence of general data extract tools for blockchains.} Although many studies provide open source data extraction tools of blockchain data, most of them can only support to extract partial blockchain data (not all the data). Moreover, most of existing tools can only fulfil specific research tasks. \textbf{(4) Absence of basic data explorations for blockchains.} Existing studies only focus on specific data analysis of blockchain data, e.g., transaction graph \cite{chen2018understanding}, contract security \cite{luu2016making}. However, the basic data explorations like statistic analysis, text analysis and data visualization are missing in most of existing tools.

To address the above challenges, we propose a blockchain data analytics framework namely XBlock-ETH to analyze Etherium data. In particular, we extract raw data consisting of 8,100,000 blocks of Ethereum. The raw data includes three types of blockchain data: \textit{blocks}, \textit{traces}, and \textit{receipts}. Since the analysis on the raw blockchain data is difficult, we process and categorize the obtained Etherium Blockchain data into six datasets: \textbf{(1) Block and Transaction}, \textbf{(2) Internal Ether Transaction}, \textbf{(3) Contract Information}, \textbf{(4) Contract Calls}, \textbf{(5) ERC20 Token Transactions}, \textbf{(6) ERC721 Token Transactions}. It is non-trivial to process the raw since it requires substantial efforts in extracting useful information from raw data and associating with six datasets. We then conduct statistic analysis on the six refined datasets. We also outlook the potential applications of XBlock-ETH, such as blockchain system analysis, smart contract analysis, and cryptocurrency analysis.

In summary, we highlight the major contributions of this paper as follows:
\begin{itemize}
    \item The XBlock-ETH data contain the comprehensive on-chain data in contrast of previous works (only cover partial Etherium data). In particular, it includes blockchain data, smart contract data, and cryptocurrency data. Moreover, the well-processed datasets can be easily used for data exploration. Furthermore, XBlock-ETH data formally released online\footnote{http://xblock.pro/dataset} has been periodically updated.
    \item The XBlock-ETH framework also offers basic statistic and exploration functions to analyze blockchain datasets.
    This paper also outlines the research opportunities brought by XBlock-ETH. In particular, we discuss the applications of XBlock-ETH in aspects of blockchain system analysis, smart contract analysis and crytocurrency analysis.
\end{itemize}

The rest of this paper is organized as follows. Section \ref{sec:back} first gives an overview of blockchain and smart contract technologies. Sections \ref{sec:rawdata}, \ref{sec:dataexploration} then present raw data acquisition from Ethereum and data exploration of six datasets. Section \ref{sec:application} discusses the applications of XBlock-ETH data. Section \ref{sec:related} surveys related work. Finally, the paper is concluded in Section \ref{sec:conc}.

\section{Background}
\label{sec:back}
\begin{figure}
    \centering
    \includegraphics[width=3.4in]{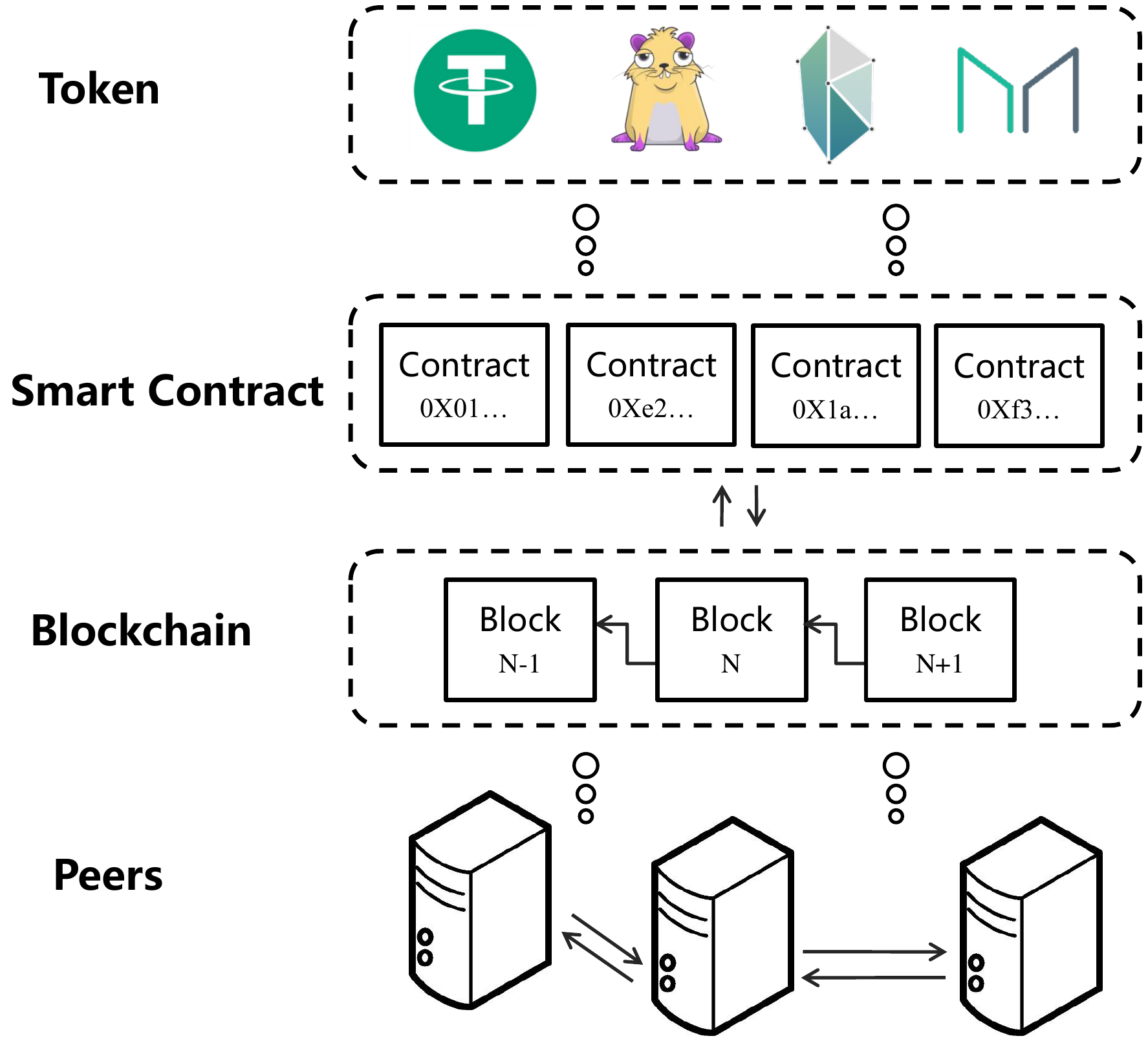}
    \caption{Overview of Ethereum Blockchain}
    \label{0_arc}
\end{figure}

Figure~\ref{0_arc} presents an overview of Ethereum blockchain, which consists of a number layers from bottom to top: peers, blockchain, smart contracts, and tokens. We next review basic concepts of each layer in Ethereum.

\subsection{Peer and Blockchain}

In a nutshell, a blockchain is essentially a chain-like data structure consisting of a number of consecutively-connected blocksblocks. The chain has been maintained by all the peers in a peer-to-peer blockchain network. In a period of time, only one block can be confirmed by the entire blockchain network through a consensus protocol. The block containing the confirmed transactions at that time and the hash value of the previous block has been generated by a \emph{peer} (a.k.a. miner). After being generated,  the block will be validated independently by the other peers. Once the block is validated and confirmed by most of peers in the blockchain network, the transactions in the block will be considered as \emph{completed}. In this way, each peer can trust the whole blockchain (a.k.a. ledger) since the transactions have been validated by all the peers. In other words, blockchain enhances trustworthiness of transactional data through duplicating computation and storage at all the peers.


Thanks to the completeness of the blockchain data in each permission-less blockchain peer, researchers can obtain the entire blockchain data via connecting a blockchain peer the blockchain network. The blockchain data that consists of all the operations done by the users and miners in the blockchain contains substantial business values. For example, the transactional records are essentially operations done by different business parties. The analysis on the blockchain data can help to understand user behaviours in a real-world economic system (e.g., money transferring). Meanwhile, there is a rapid growth of blockchain data, especially in Bitcoin and Ethereum, with the proliferation of blockchain users and transactions. The analysis on blockchain data can be also beneficial to predict the economic trend.

\subsection{Smart Contract}

Smart contract that was proposed even earlier than blockchain \cite{szabo1997idea} is a promising technology to reshape the modern industry. Blockchain-based smart contracts are essentially computer programs, in which the execution states are stored on top of blockchain. The blockchain transactions are the messages representing the deployment or invocations of smart contracts. Therefore, blockchain guarantees the trustworthiness of smart contracts.

The incumbent blockchain systems have enable smart contracts. For example, Bitcoin enables users to run a simple script program during the execution of transactions. This script can be regarded as a simple blockchain-based smart contract. However, the Bitcoin script is not Turing-complete so that it cannot enables complex logic expressions in the contract. In contrast, Ethereum enables Turing-complete smart contracts. In Ethereum, smart contract is executed in the environment called Ethereum Virtual Machine (EVM). EVM reads and writes the states (stored in the key-value like database) as the actions defined in a smart contract. During the contract execution, a miner uses ``Gas'' as a unit to evaluate the consumption of one smart contract. After running the contract , the contract user is charged by the ``\texttt{GasUsed}'' and ``\texttt{GasPrice}''. The more ``\texttt{GasPrice}'' that the users promise to pay for the miner, the faster the contract executes. After the transactions (i.e., operations) are done, EVM will generate a hash value of the state and record it into the blockchain. Therefore, we can learn from Figure~\ref{0_arc} that smart contracts on Ethereum are not directly stored on blockchain. They are essentially stored in the states that have been operated by the blockchain.

\subsection{Tokens and clients}
It is worth mentioning that Ethereum has two standard token protocols (a.k.a. templates) of smart contracts \cite{buterin2015erc20,entriken2018erc}. These token protocols define the standard variables, functions, and interfaces in the smart contract. With the protocols, users can issue tokens (or so-called cryptocurrencies) based on smart contracts on top of Ethereum. There are four typical tokens USDT\footnote{https://tether.to/}, Cryptokitties\cite{kharif4cryptokitties}, Kyber\cite{luukybernetwork}, MarkerDAO\footnote{https://makerdao.com/} as shown in Figure~\ref{0_arc} (i.e., the top layer). For an example, a user can publish an ERC20 contract on Ethereum issuing tokens to others. After that, any other users (even contracts) can receive or send the token without a centralized authority (e.g., stock exchange). The standard token protocols greatly enrich the ecosystem of Ethereum so as to make Ethereum become a more flexible financial system. In Section~\ref{Dataset 5: ERC20 Token Trasnaction} and \ref{Dataset 6: ERC721 Token Trasnaction}, we will explore the data of tokens in Ethereum.

Ethereum allows that any computer programs can join into the network if they meet the requirement of the protocol just like P2P protocols (e.g., BitTorrent). As a result, there are a number of diverse Ethereum clients that can validate the blocks and transactions. Among most of Ethereum clients, Go-Ethereum (Geth) and Parity have been the most widely used according to the statistic from Ether nodes\footnote{https://ethernodes.org}. Both of them provide JSON-RPC interfaces for users to interact with Ethereum blockchain. Through the JSON-RPC interfaces, user can obtain the blockchain data from Ethereum. Geth has been generally used in many previous studies while the interfaces designed in Geth is not suitable for data acquisition. Even though many researchers attempted to modify source codes of Geth to obtain the detailed run-time data, the whole procedure of the code modification is time consuming and complex. In addition, the obtained data is not absolutely accurate in some cases. Different from Geth, Parity better designs the interfaces so that it can obtain the index of each block corresponding to each piece of the data that we need. The details on data acquisition of blockchain data will be described in Section~\ref{sec:rawdata}.

\section{Raw data extraction from Ethereum} 
\label{sec:rawdata}
\begin{figure}
    \centering
    \includegraphics[width=3.6in]{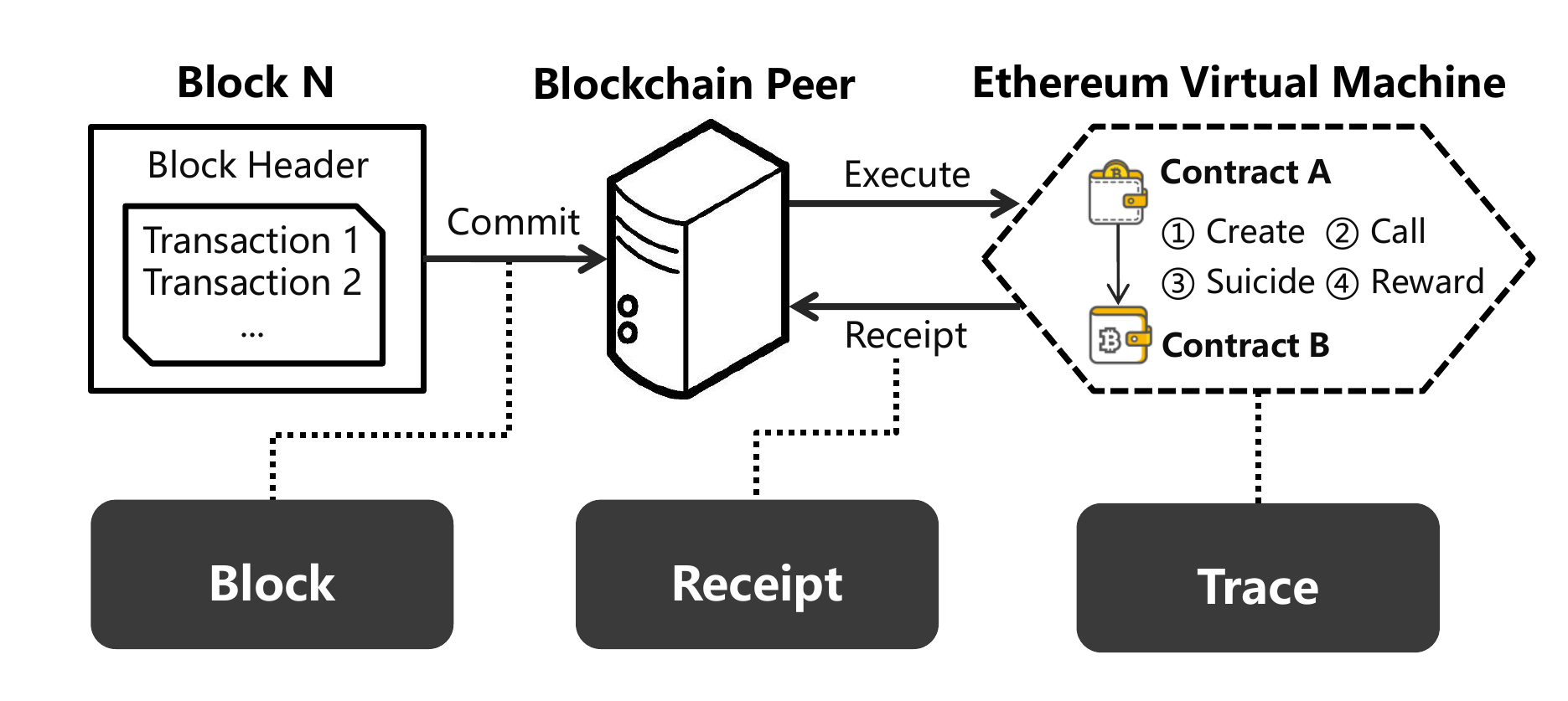}
    \caption{Raw data collection during Ethereum transaction flow}
    \label{0_rawdata}
\end{figure}

This section describes the procedure how the raw data was obtained from Ethereum blockchain. Figure~\ref{0_rawdata} illustrates the typical Ethereum transaction execution flow from Block $N$ to EVM through 
Blockchain peer. During this procedure, we collect the three types of blockchain raw data: Block, Receipt and Trace. We next describe the details on the composition and acquisition of each kind of raw data.

\subsection{Block}
Block data is directly stored in Ethereum blockchain. Each block consists of two components:
\begin{itemize}
    \item \textbf{Block Header: } Block header is the basic information of a block, including the miner's address, timestamp, gas limit, etc.
    \item \textbf{Block Transactions: } Block transactions constructs the body of the block. Each transaction consists of the fields: From, To, Value, Input, etc. If the transaction is used to deploy a contract, the \emph{To} field is ``\texttt{null}'' in the block transaction.
\end{itemize}
Almost all the Ethereum clients including Geth and Parity offer the interfaces to query the blocks. For example, ``\texttt{eth\_getBlock}'' is available in both Geth and Parity with the similar efficiency.

However, we can only obtain little information about the blockchain users through analyzing the block data. This is because the input of block transaction only represents operations to EVM in the contract deployment phase while the contract code will be stored only at the end of the transaction execution and it is not the same as the input of the transaction. Thus, we cannot obtain the exact contract code in the block transaction. Meanwhile, in the contract invocation phase, we cannot know whether the transaction is executed successfully or what kinds of error thrown during the transaction execution since sometimes a contract will send messages or cryptocurrencies to other contracts.

\subsection{Trace}
Trace data is essentially the detailed run-time data that was generated in EVM (e.g., internal contract calls, transferring money from the contract to a person). Trace data cannot be directly obtained or observed from the block data, but can be recorded during the contract execution. In this paper, trace data is referred to the data that cannot be obtained before or after the transaction execution, but only appears during the execution. Trace data includes the following types:

\begin{itemize}
    \item \textbf{Create} is the trace including the creator, code, and initial balance when a smart contract is deployed. The creator of a contract can be a person or another smart contract.
    \item \textbf{Call} occurs when money or messages are transferred through different Ethereum addresses. Contract call or Ether transferring is shown as a ``\texttt{Call}'' trace.
    \item \textbf{Suicide} is the trace that smart contract ``\texttt{suicide}'' deletes its code, and refunds the value to a specific account.
    \item \textbf{Reward} is the trace that miners get the Ether reward when they mine a block. The reward value varies depending on the contribution of the miners.
\end{itemize}

In Geth, the interface of trace is ``\texttt{debug\_traceTransaction}''. However, this interface returns all the operations during the transaction, resulting in large resource consumption and low efficiency. Thus, many previous studies attempt to modify the source codes of Geth to obtain the detailed run-time data, while this procedure is extremely time consuming. 

In this paper, we adopt ``\texttt{parity\_trace}'' in Parity to obtain the trace data. This interface is provided and maintained by the official developer so that the correctness is guaranteed in contrast to Geth. Meanwhile it also provides enough information that we need, such as the basic trace types and errors. Moreover, another advantage of Parity is the updating convenience as the data is indexed by blocks.

\subsection{Receipt}
After the transaction is executed, some of the Ethereum states have been changed (e.g., the balance of the account in a token contract). Then the clients need to know what have been changed. To reduce the query overhead of clients, many contracts leave a kind of outputs called ``\textit{Event}'' in the execution. For example, a standard token contract will output a ``\texttt{Transfer(from,to,value)}'' event to let the clients know what happens during the execution. This kind of outputs is an one-way output, as it is just written in the receipt of the transaction, and can be read by external clients or persons but cannot be read by internal EVMs.

Section \ref{sec:dataexploration} will then give the statistics of Ethereum data. In particular, there are over 100,000 kinds of cryptocurrencies using smart contracts on Ethereum. As for these token contracts, the receipt data is the important source to learn about the holders, owners, and the user behaviors. Thus, it is necessary to obtain receipt data.

Both Geth and Parity provide the interfaces to get the transaction receipts. The main difference between Geth and Parity interfaces lies in the query index of the receipts. In particular, the receipt of the interface of Geth is ``\texttt{eth\_getTransactionReceipt}'' that is indexed by the transaction hash, while the interface of Parity is ``\texttt{parity\_getBlockReceipts}'' that is indexed by block number. In this way, Parity is much more efficient than Geth since it can return a batch of receipts in one query.

In summary, there are three kind of raw datasets that can be obtained in Ethereum: block, trace, and receipt. Because of the massive volume and redundant information of the raw data, data procession is necessary to simplify data representation and fasten data analysis for the further study. After compression, the size of the data is about 313 GBytes. 

\section{Data exploration of Ethereum} 
\label{sec:dataexploration}

\begin{figure}
    \centering
    \includegraphics[width=3.4in]{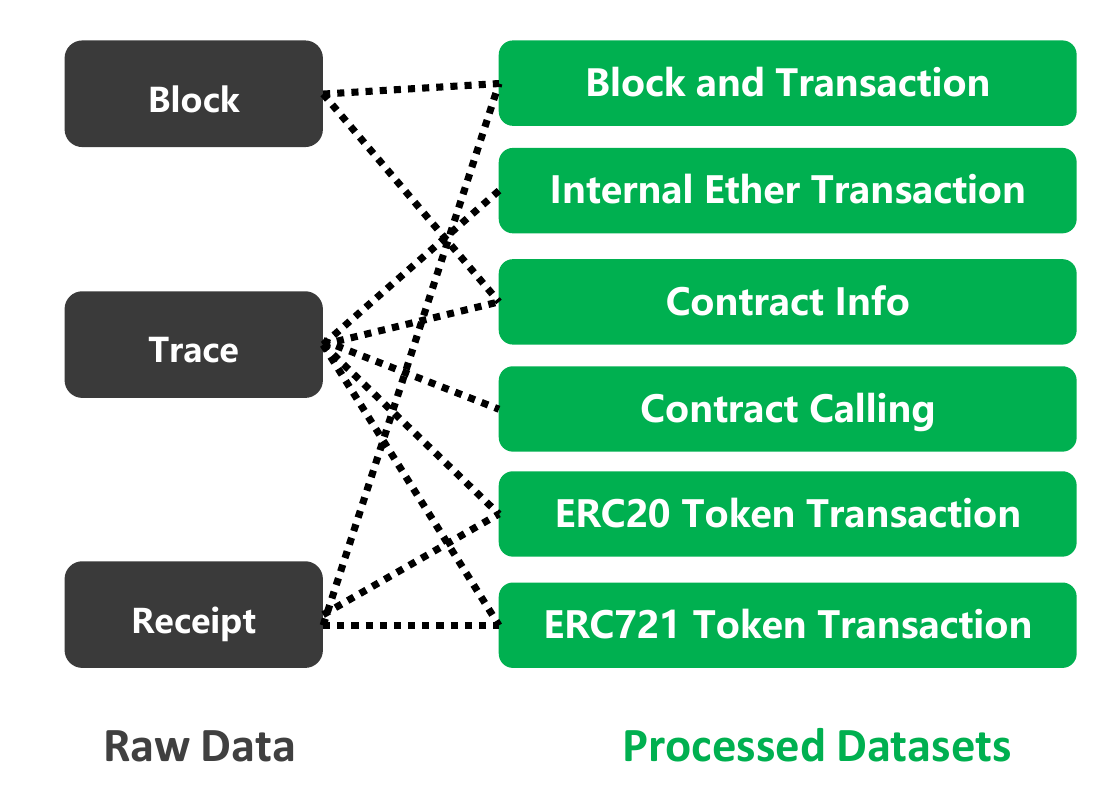}
    \caption{Mapping from raw data to datasets}
    \label{0_washdata}
\end{figure}

In this section, we process the obtained raw data from Ethereum and divide it into six datasets: (1) Block and Transaction, (2) Internal Ether Transaction, (3) Contract Info, (4) Contract Call, (5) ERC20 Token Transaction, (6) ERC721 Token Transaction. The relationship from the raw data to the processed datasets  is shown in Figure~\ref{0_washdata}. We can easily observe that the trace data has been the most widely used in the data process. This section will introduce how the datasets are generated, with statistics and observations.

\subsection{Dataset 1: Block and Transaction} \label{Dataset 1: Block and Transaction}

\begin{table}[]
    \centering
    \caption{Statistics of Dataset 1}
    \begin{tabular}{|l|r|}
    \hline
    \textbf{Statistics} &  \textbf{Values} \\ \hline\hline
    No. of Blocks     &   8,100,000 \\ \hline
    No. of Transactions &  491,562,222 \\ \hline
    No. of Miner Addresses & 5,122\\  \hline
    Mean of Transaction Counts per Block & 60.68  \\ \hline
    Mean of Block Time & 15.33 seconds \\ \hline
    Mean of Block Size & 11,457 bytes \\ \hline
    \end{tabular}
    \label{Statistics of Dataset 1}
\end{table}

\renewcommand\subfigcapskip{-0.5ex}
\begin{figure}
\centering 
\subfigure[Word Cloud of Miners' Text]{
\centering
\includegraphics[width=4.1cm]{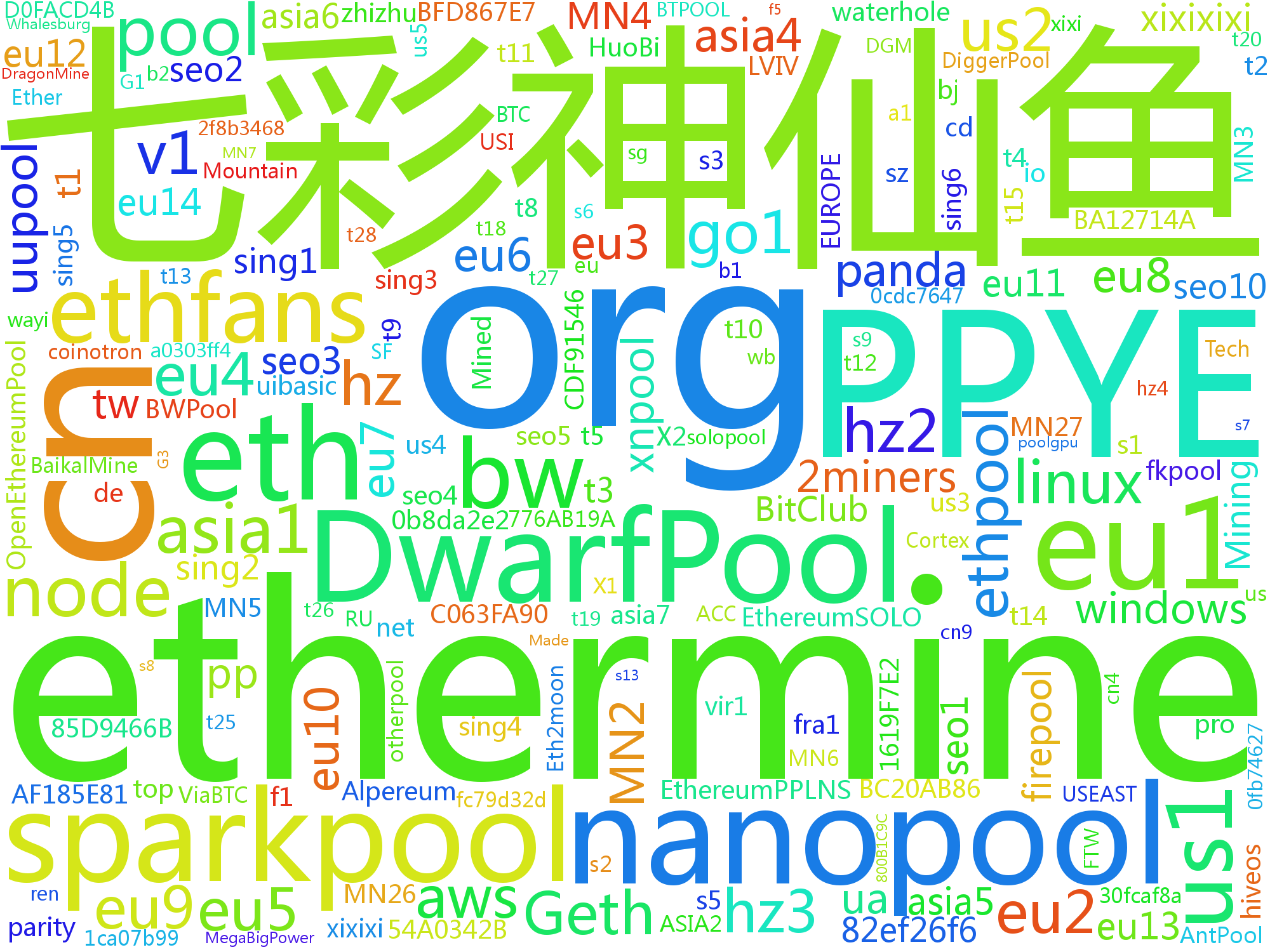}\label{1_extraCloud}
} 
\subfigure[Transaction Count]{
\centering
\includegraphics[width=4.1cm]{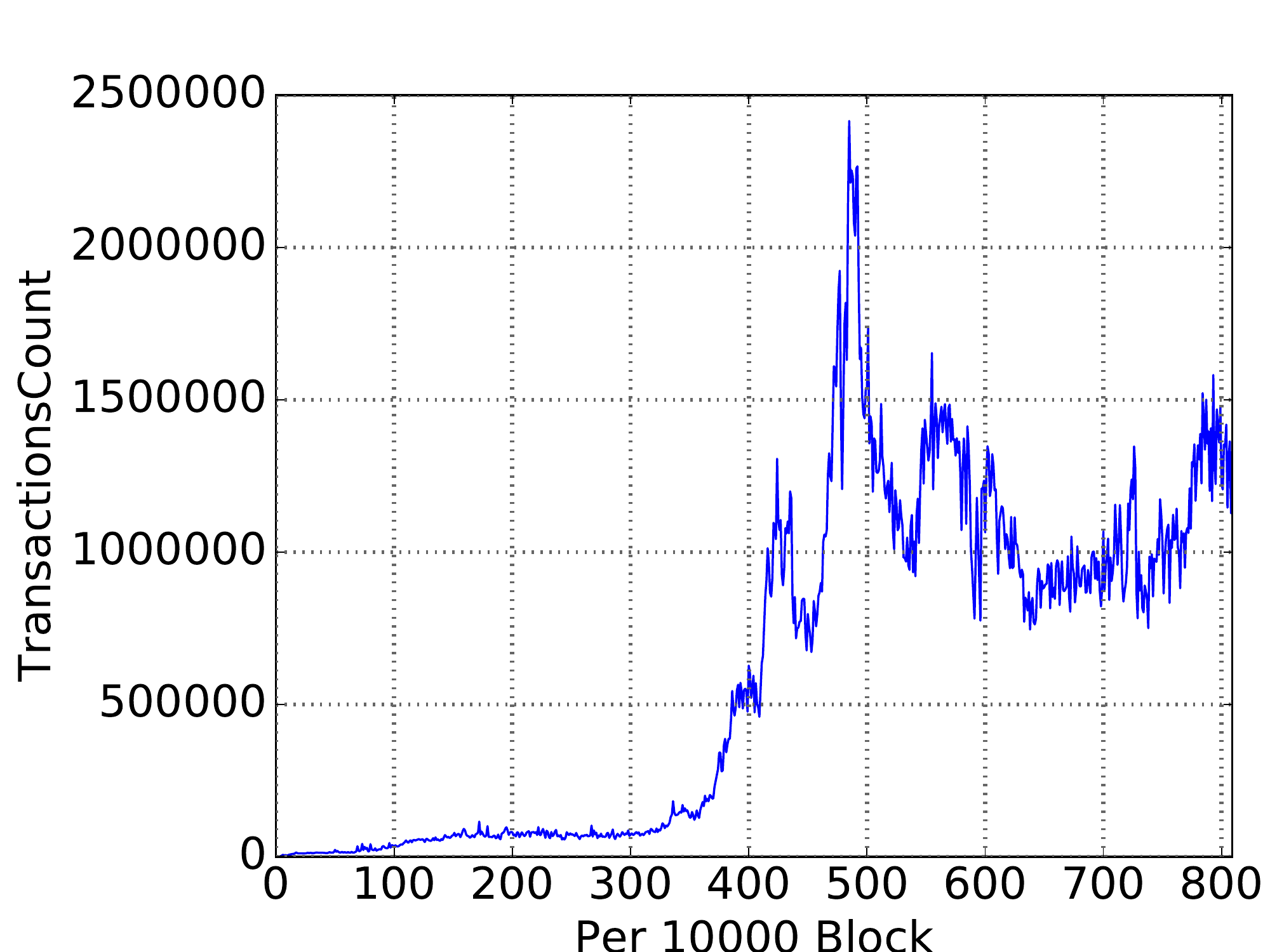}\label{1_txcount}
}
\subfigure[Macro view of GasPrice]{
\centering
\includegraphics[width=4.1cm]{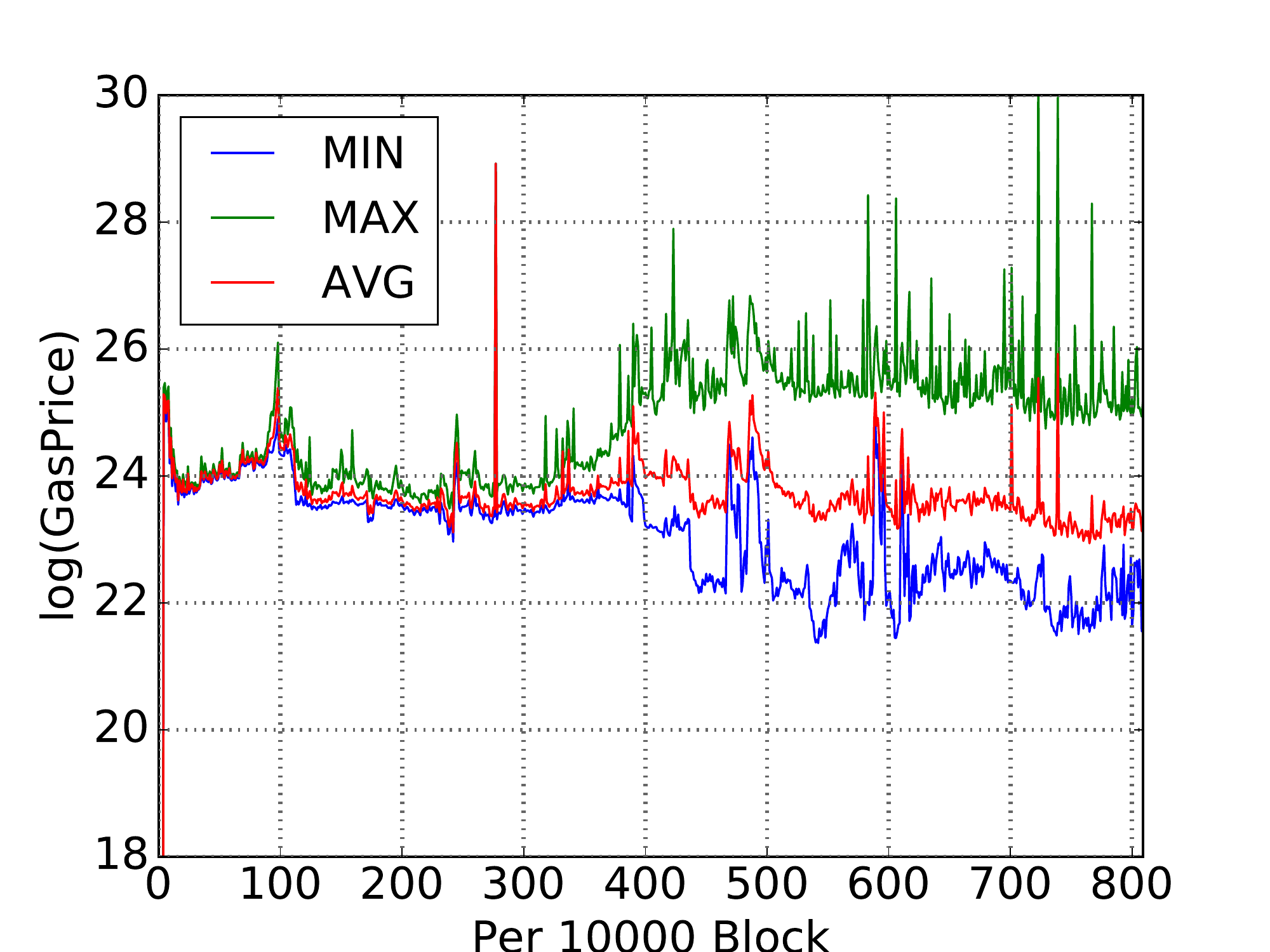}\label{1_gasprice}
} 
\subfigure[Micro view of GasPrice]{
\centering
\includegraphics[width=4.1cm]{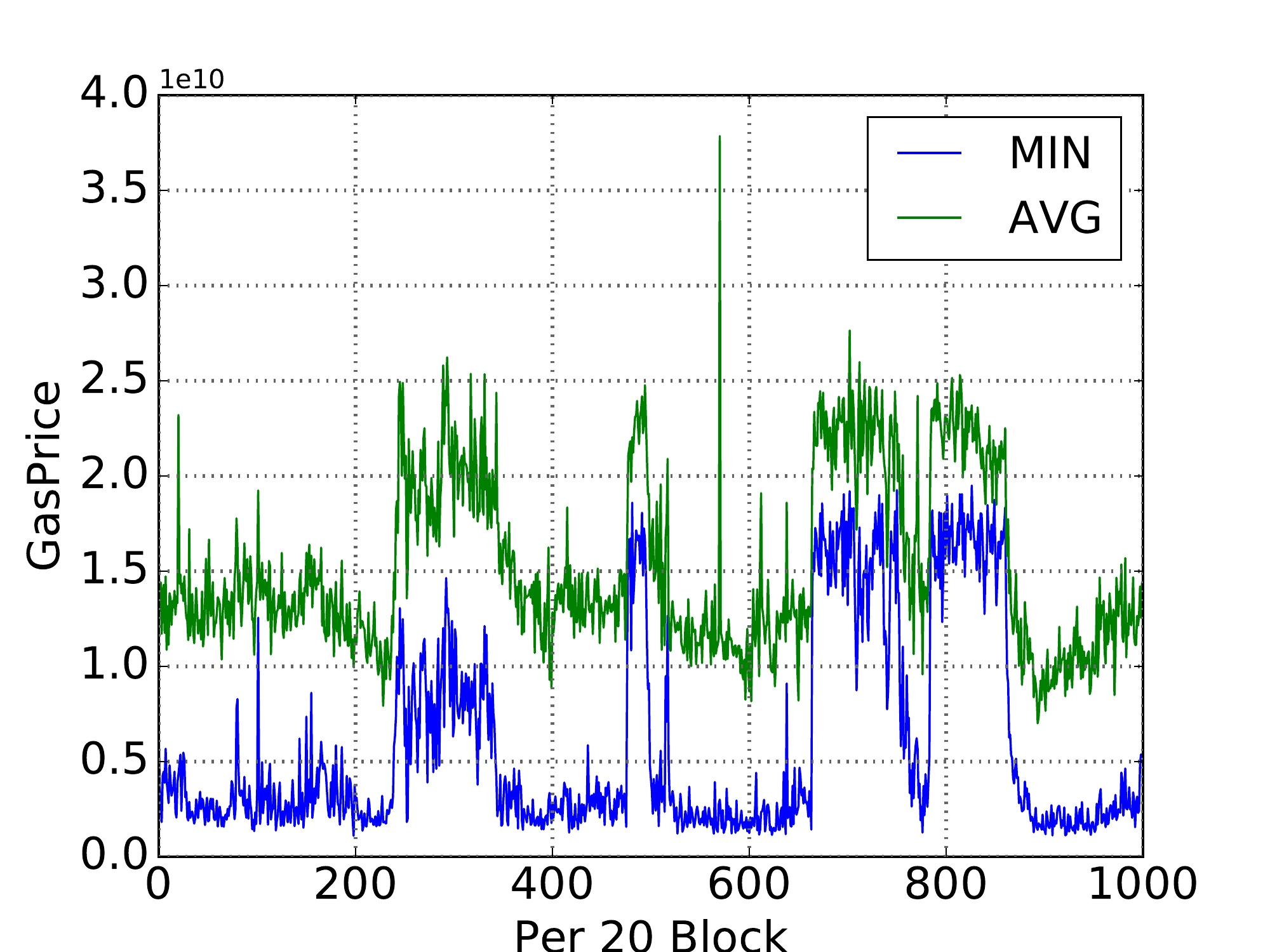}\label{1_gaspriceOneday}
}

\caption{Visualization of Dataset 1} \label{cost}
\end{figure}

To investigate the basic statistics of Ethereum, we extract the information about the blocks and the transactions inside the blocks. In particular, there are 8,100,000 blocks and 491,562,222 transactions  generated from the block data. For each block, we also obtain the statistic values of the ``\texttt{gasPrice}'': minimum, average, and maximum. Meanwhile, corresponding to the hash of each transaction, the fields of ``\texttt{minerReward}'', ``\texttt{gasUsed}'' and ``\texttt{error}'' are extracted from the receipt and trace.

Regarding the miners of the Ethereum blockchain, there are 5,122 unique addresses of miners as shown in Table~\ref{Statistics of Dataset 1}. It implies that there are no more than 5,122 peers that serve as miners since one peer may own more than one addresses. Meanwhile, each miner has the right to write extra texts in the block. So, we also use the word cloud to analyze the texts of miners. Figure~\ref{1_extraCloud} shows the visualization of the texts of the word cloud. The results show that there are  texts left by the mining pool, since most miners are in the mining pool and they have left their names in the blocks to promote their mining capability. 

As shown in Table~\ref{Statistics of Dataset 1}, the mean of transaction counts per block is 60.68, and the block time is 15.33 seconds. In other words, the average throughput of Ethereum is about 4 transactions per second. Even when most of the network is active, as shown at 4,900,000 blocks in Figure~\ref{1_txcount}, the throughput is about 16.7 transactions per second. This result implies that Ethereum still has a long way to go to support real-time Internet applications.

In Ethereum, a miner has a higher priority to package the transactions with higher ``\texttt{gasPrice}'' into the block. The visualization of ``\texttt{gasPrice}'' is shown in Figures~\ref{1_gasprice} and~\ref{1_gaspriceOneday}. In a macro view, the ``\texttt{gasPrice}'' is gradually decreasing with the development of the Ethereum community, except for  several peaks caused by extremely frequent transaction when the network is congested. In a micro view, we extract the time from 8,000,000 to 8,020,000 blocks and find that such fluctuations of ``\texttt{gasPrice}'' can be observed by the tidal law. This observation implies that the fluctuations of ``\texttt{gasPrice}'' can potentially be predicted.

\subsection{Dataset 2: Internal Ether Transaction} \label{Dataset 2: Internal Ether Transaction}

Ether is the native cryptocurrency of Ethereum. The transactions of Ether not only happen in the transactions recorded in the block, but also occur during the smart contract execution. For example, if someone asks a smart contract to send 10 Ethers to another one, the Ether transaction from the contract will not be observed in the block. In some blockchain explorers such as Etherscan\footnote{http://etherscan.io}, this kind of transactions is also called ``\emph{Internal Transaction}''. To investigate all the Ether transactions, we process the block and trace data to conduct the internal Ether transaction dataset. As shown in Table~\ref{Statistics of Dataset 2}, 329,020,672 Ether transactions which occur among 54,720,018 addresses are collected. 

\begin{figure}
\centering 
\subfigure[Ether Transferred Amount]{
\centering
\includegraphics[width=4.1cm]{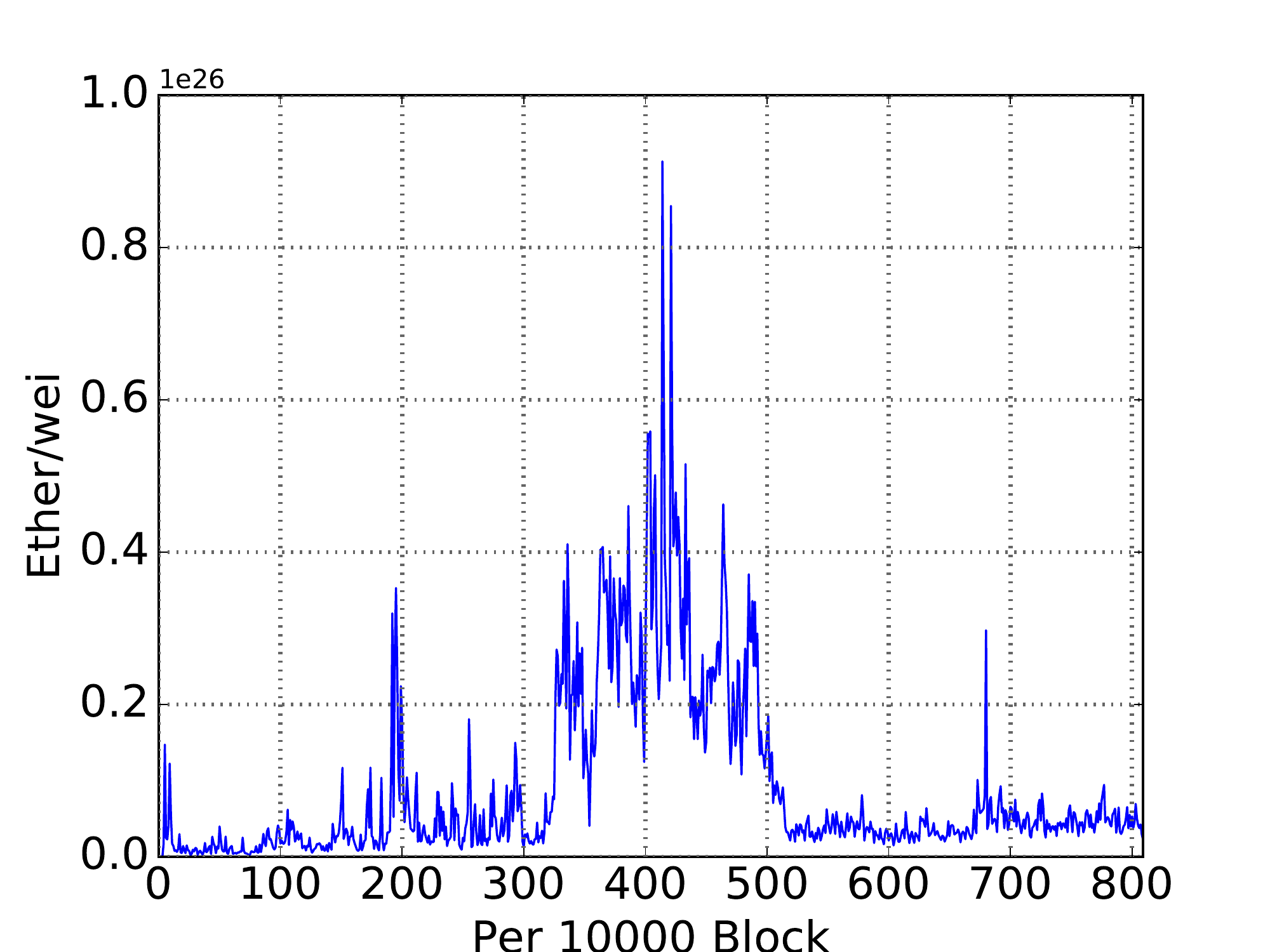} \label{2_ethertime}
} 
\subfigure[Ether Transaction Distribution]{
\centering
\includegraphics[width=4.1cm]{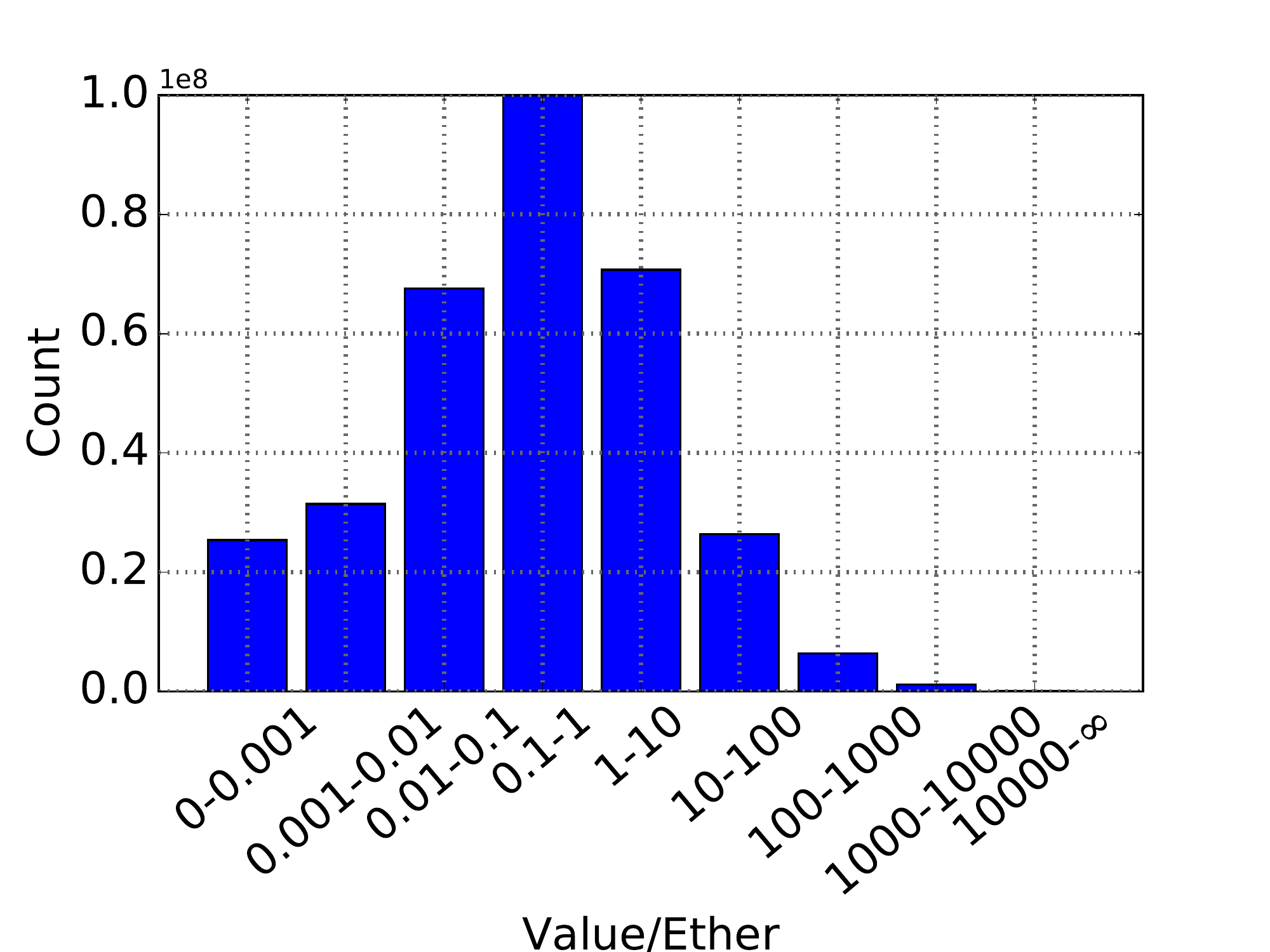} \label{2_ethervalue}
}

\caption{Visualization of Dataset 2} \label{cost}
\end{figure}

The values of Ether have a large variance, as the maximum is 11,901,464.24 Ethers (about 2 billions dollars now) but the mean is only 22.30 Ethers. Figure~\ref{2_ethertime} presents statistics on the total transaction amount of every 10,000 blocks. It is shown that the most active time for Ether transaction is the time during 4,000,000 to 4,300,000 blocks, matching with the most active time of Initial Coin Offering (ICO). Regarding the Ether distribution as shown in Figure~\ref{2_ethervalue}, we find that most of Ether transactions fall in the range from 0.1 Ether to 1 Ether, indicating that most of transactions only transfer small amounts of Ethers.

\begin{table}[]
    \centering
    \caption{Statistics of Dataset 2}
    \begin{tabular}{|l|r|}
    \hline
    \textbf{Statistics} &  \textbf{Values} \\ \hline\hline
    No. of Ether Transactions     &   329,020,692 \\ \hline    
    No. of Addresses     &  54,720,018 \\ \hline               
    Mean of Amount of Ethers     &  22.30 \\ \hline
    Maximum of Amount of Ether     &  11,901,464.24\\ \hline
    \end{tabular}
    \label{Statistics of Dataset 2}
\end{table}

\renewcommand\subfigcapskip{-0.5ex}

\subsection{Dataset 3: Contract Info} \label{Dataset 3: Contract Info}

\begin{table}[]
    \centering
    \caption{Statistics of Dataset 3} \label{Statistics of Dataset 3}
    \begin{tabular}{|l|r|}
    \hline
    \textbf{Statistics} &  \textbf{Values} \\ \hline\hline
    No. of Created Contracts     &   16,609,273  \\ \hline    
    No. of Creator Addresses     &   133,484\\ \hline  
    No. of Deleted Contracts     &   5,564,823  \\ \hline    
    No. of Refunded Addresses     &  19,133,481\\ \hline
    Mean of Contract Hex Code Size     &  958.20  \\ \hline
    \end{tabular}
    \label{Statistics of Dataset 2}
\end{table}

\renewcommand\subfigcapskip{-0.5ex}
\begin{figure}
\centering 
\subfigure[Contract size distribution]{
\centering
\includegraphics[width=4.1cm]{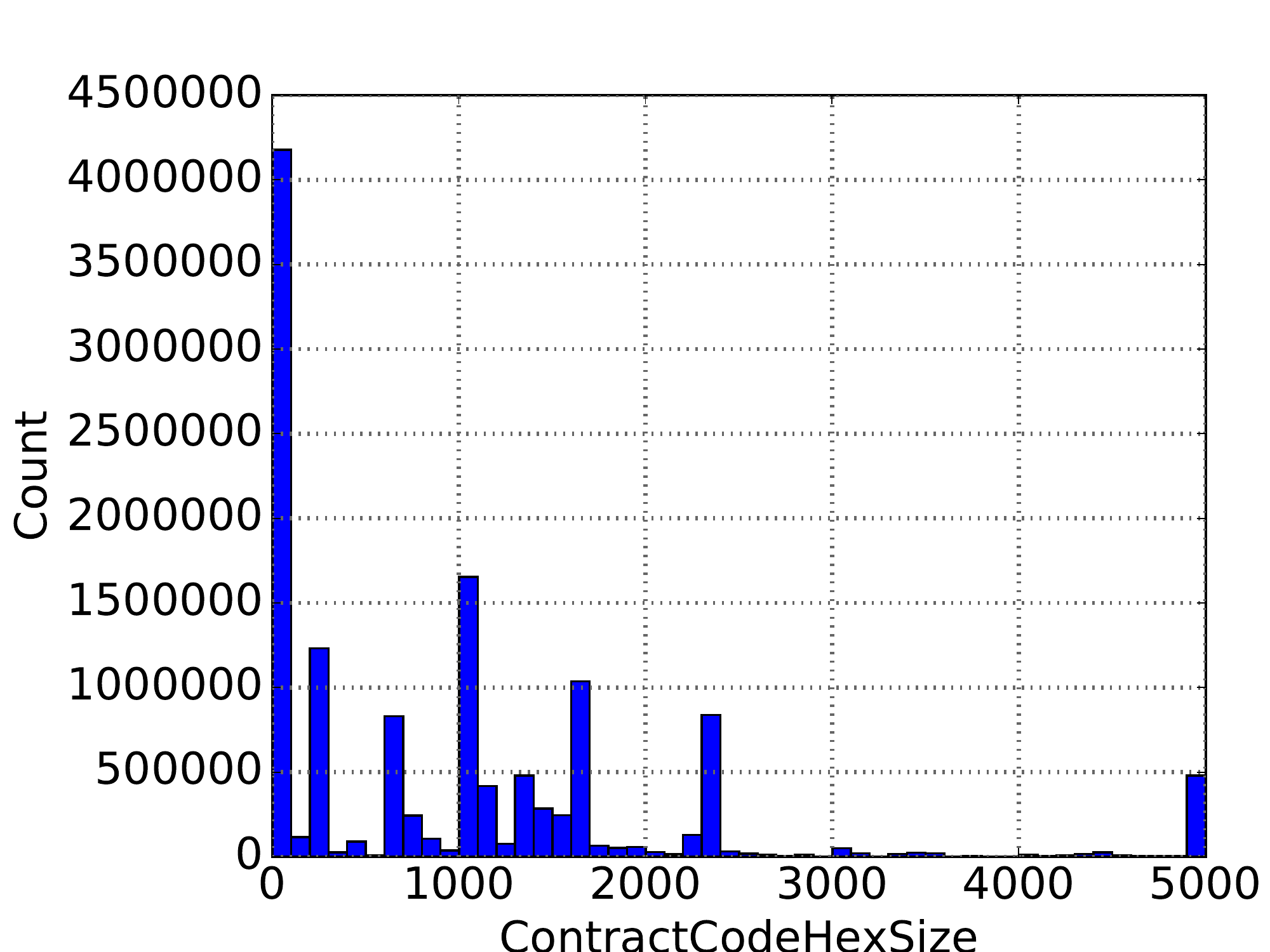}\label{3_contractsize}
} 
\subfigure[Count of created contracts ]{
\centering
\includegraphics[width=4.1cm]{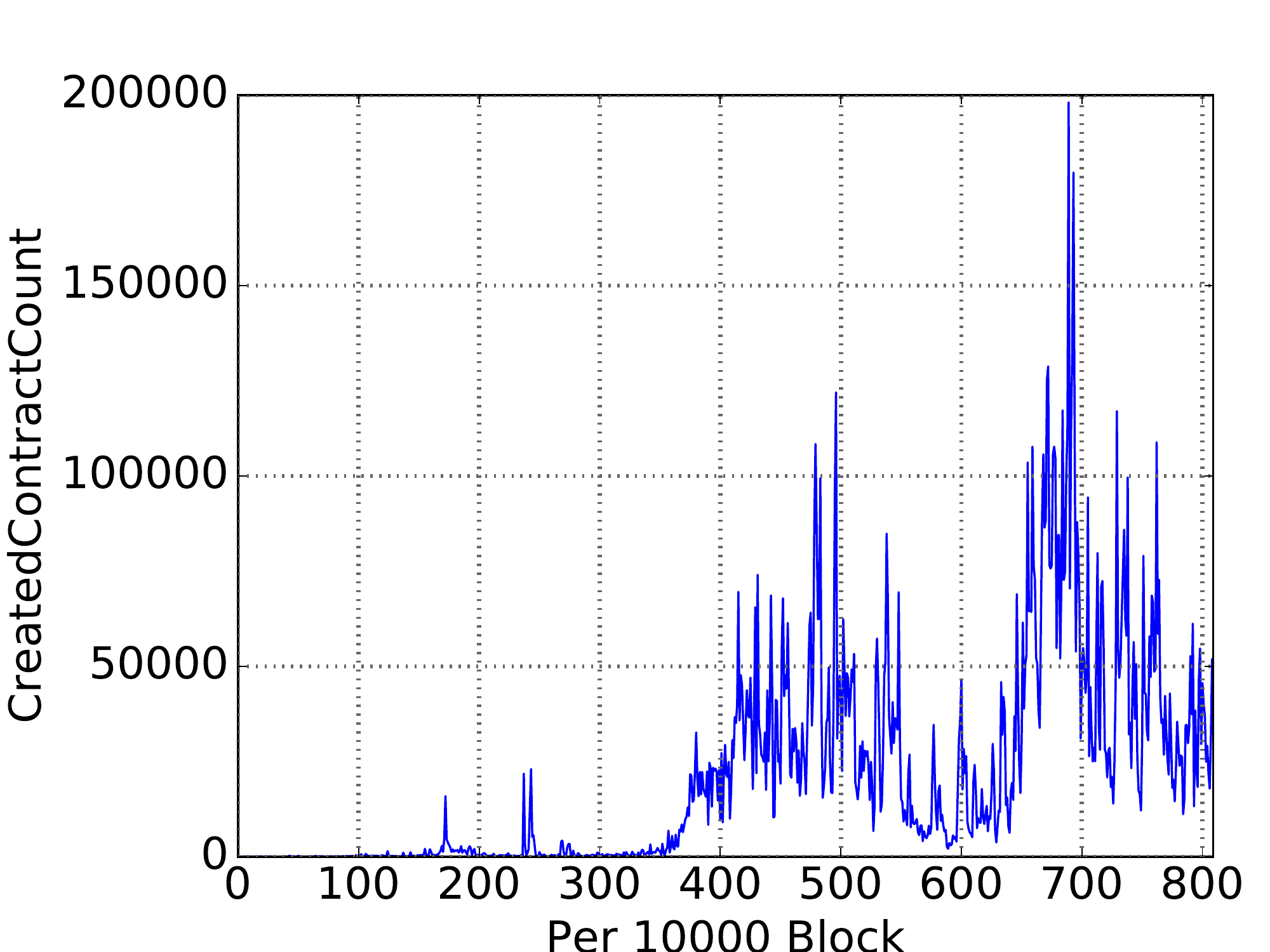}\label{3_contracttime}
}

\caption{Visualization of Dataset 3} \label{cost}
\end{figure}

Ethereum can be considered as a platform for smart contracts. To investigate all the smart contracts on Ethereum, we process the trace data to get the basic information of smart contracts, including the creator, created-time, initial value, contract code, creation code. Some smart contracts can be deleted and refund Ethers to someone if they set a ``SUICIDE'' operation code inside a function. Therefore, we can observe the actions of contract deletions. According to the statistics in Table~\ref{Statistics of Dataset 3}, there are 16,609,273 smart contracts created by 133,484 addresses. It implies that there should be a number of users who create multiple contracts.

An abnormal phenomenon observed from Table~\ref{Statistics of Dataset 3} is that 5,564,823 contracts are deleted while they refund the Ether balance to 19,133,481 addresses. Generally, a smart contract will not refunds Ethers to multiple addresses during deletion. The reason behind this abnormal phenomenon is that Ethereum has suffered from a Denial of Service (DoS) attacks, in which attackers use a vulnerability of the price of ``SUICIDE'' to create accounts in Ethereum. Before the vulnerability is fixed, a great amount of contracts are deleted to direct to empty address, leading to many Ethereum peers shutting down as indicated in previous work \cite{chen2017adaptive}.

Regarding the contract code, we translate the bytecode into hexadecimal code. Figure~\ref{3_contractsize} gives the statistics of contract size. Particularly, the mean of contract size is 958.20, indicating that the smart contracts take up little space of storage. The contract size distribution also implies that the sizes of most contracts have focused on some clusters. This indicates that many smart contracts may look similar. This similarity will be further investigate in Dataset 4. Figure~\ref{3_contracttime} presents the count of created contracts. It is shown in Figure~\ref{3_contracttime} that the number of new smart contracts is increasing, especially at the time after the concept of ``ICO''\cite{howell2018initial} comes out. 

\subsection{Dataset 4: Contract Call} \label{Dataset 4: Contract Call}

In EVM, a smart contract can call another one to invoke some codes or functions. To investigate the calls among the Ethereum contracts (which are represented as addresses), we extract Contract Calls in the execution from the trace dataset. The contract call dataset includes the caller, called address, calling function. As shown in Table~\ref{Statistics of Dataset 4}, it consists of 1,148,572,009 Contract Calls, among which 639,336,722 contain input codes and 169,463,261 contain errors.

\begin{figure}
\centering 
\subfigure[Count of Contract Call]{
\centering
\includegraphics[width=4.1cm]{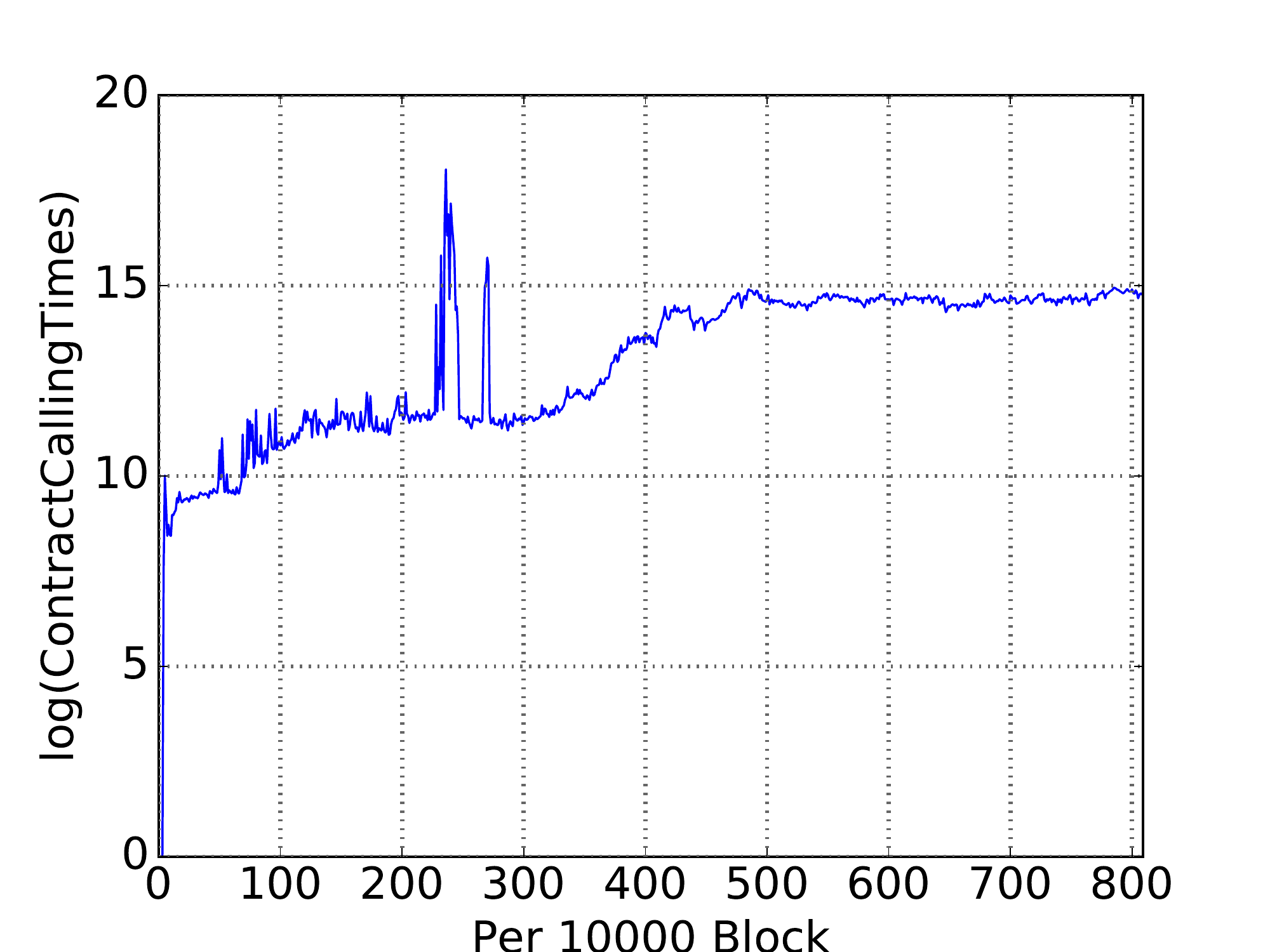}\label{4_calltime}
} 
\subfigure[Call Type Distribution]{
\centering
\includegraphics[width=4.1cm]{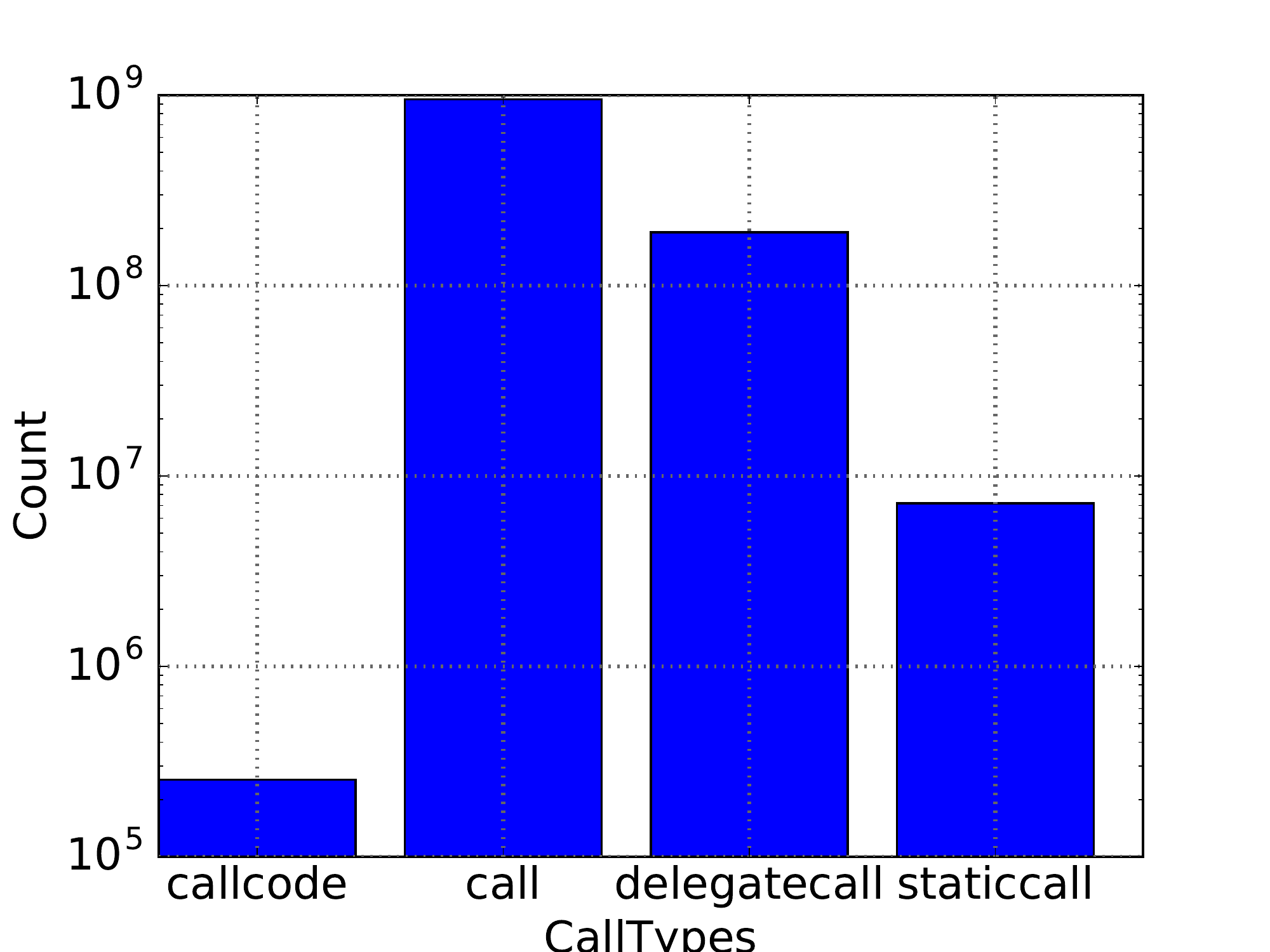}\label{4_calltype}
}
\subfigure[Count of Contract Error]{
\centering
\includegraphics[width=4.1cm]{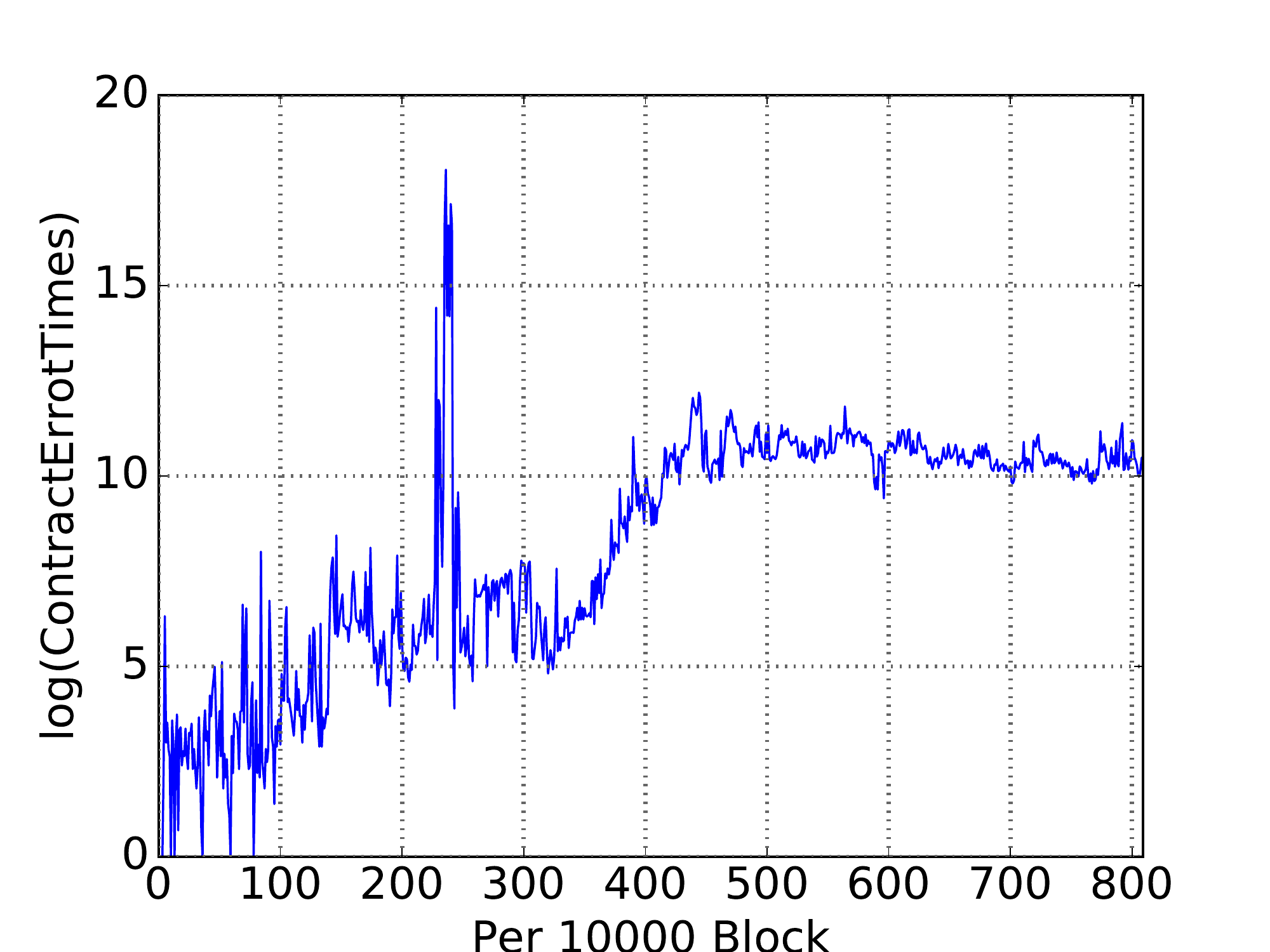}\label{4_errortime}
} 
\subfigure[Error Type Distribution]{
\centering
\includegraphics[width=4.1cm]{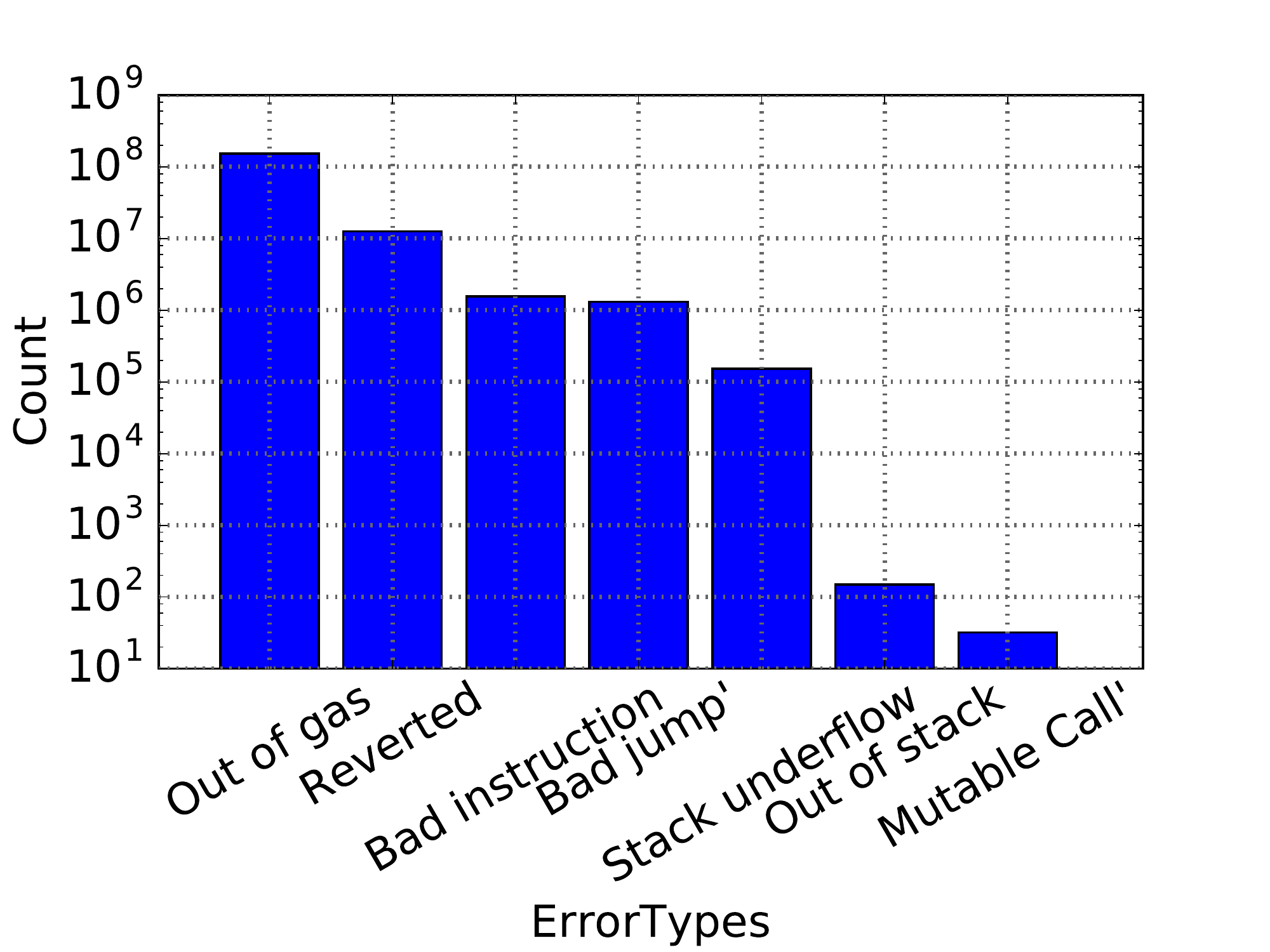}\label{4_errortype}
}
\subfigure[Calling Count of Top 10 Contract Function]{
\centering
\includegraphics[width=5cm]{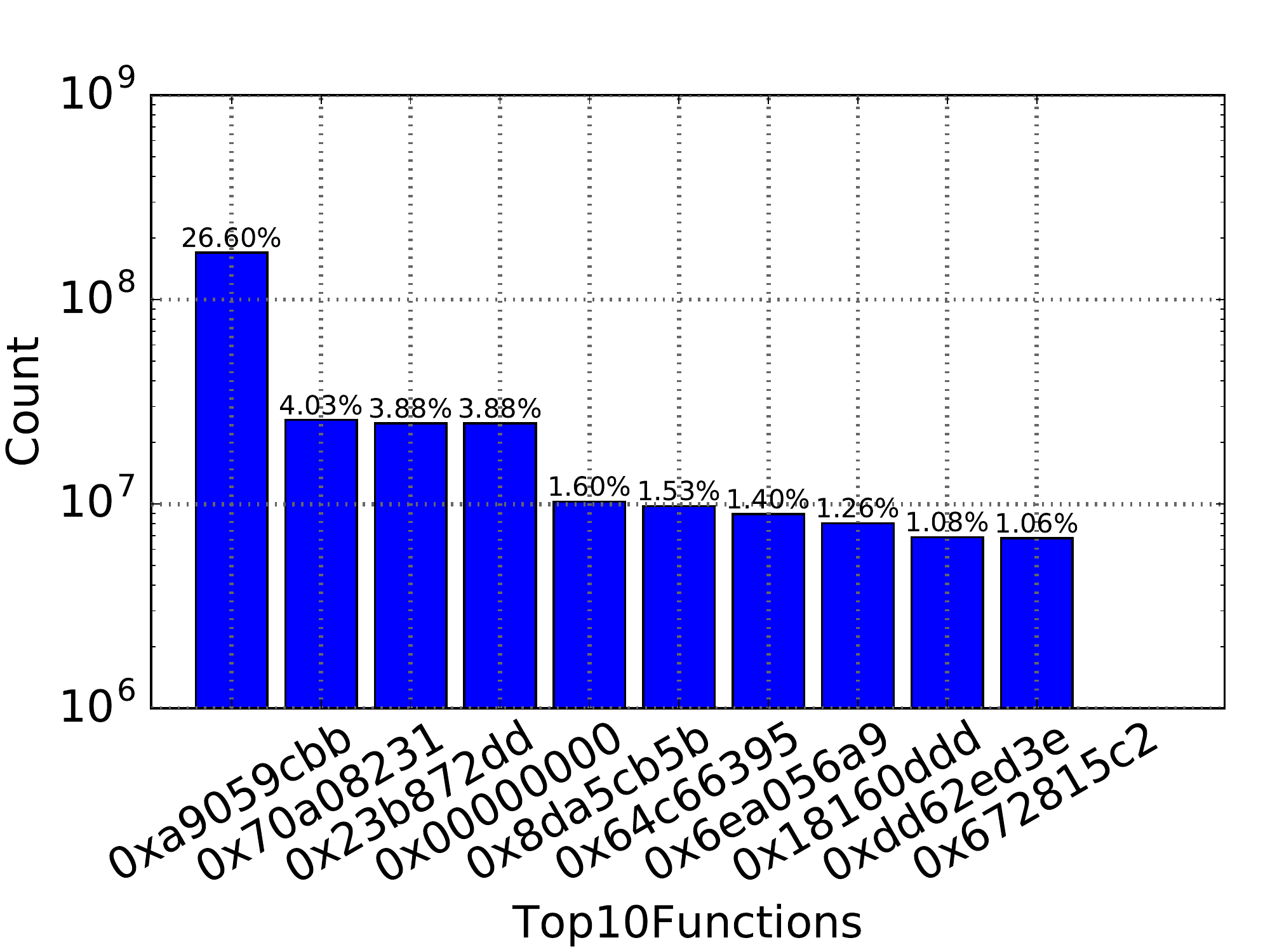}\label{4_funccount}
}
\caption{Visualization of Dataset 4} \label{Visualization of Dataset 4}
\end{figure}

Figure~\ref{Visualization of Dataset 4} gives the visualization of Contract Calls. In particular, Figure~\ref{4_calltime} and Figure~\ref{4_errortime} show that, during the time from 2,300,000 to 2,460,000 blocks, contract calls and errors occur very frequently. This is caused by the DoS attacks mentioned in the above subsection, as the attackers invoked a large number of contracts in batches and some of them throw errors. Figure~\ref{4_calltype} gives the distribution of call types. In particular, Figure~\ref{4_calltype} shows that most of developers prefer to use ``\texttt{call}'' and ``\texttt{delegatecall}'' rather than ``\texttt{staticcall}'' and ``\texttt{callcode}'', since the logic of ``\texttt{call}'' and ``\texttt{delegatecall}'' is clearer and more practical than other two calls. Figure~\ref{4_errortype} shows the error types during calling contract, indicating that most of errors are caused by ``Out of gas'', which is mainly resulted from the wrong settings of message senders. The second most common error is ``\texttt{Reverted}'', which is a manually-thrown exception by the developers. Moreover, other errors such as ``Bad instruction'' and ``Bad jump destination'' are often caused by the contract codes themselves.

Generally, the compiler of smart contracts will use the hash value of function name and parameters as the entry of the function. In other words, in Ethereum smart contracts, the identical function in source code will have the identical entry in the complied contract code. We then count the calling contract functions to see what functions are the most common ones. The distribution of top-10 functions is shown in Figure~\ref{4_funccount}. The results show that most of the calling functions concentrated on some types of them. For example, top-10 functions have occupied 46.32\% of the contract calls. Moreover, after verifying the hash values of functions with the open-source contracts, we obtain the functions in source code. We then have the top-3 functions: ``\texttt{transfer(address,uint256)}'', ``\texttt{balanceOf(address)}'' and ``\texttt{transferFrom(address,address,uint256)}''. This result implies that the most common contract calls are about tokens and there might be a great similarity among the contracts due to the similar calls.

\begin{table}[]
    \centering
    \caption{Statistics of Dataset 4}
    \label{Statistics of Dataset 4}
    \begin{tabular}{|l|r|}
    \hline
    \textbf{Statistics} &  \textbf{Values} \\ \hline\hline
    No. of Contract Calls     &   1,148,572,009  \\ \hline    
    No. of Calls with Inputs     &  639,336,722 \\ \hline  
    No. of Calls with Errors    &   169,463,261   \\ \hline
    \end{tabular}
\end{table}

\renewcommand\subfigcapskip{-0.5ex}

\subsection{Dataset 5: ERC20 Token Trasnaction} \label{Dataset 5: ERC20 Token Trasnaction}

From the above analysis, we observe that the most active smart contracts on Ethereum now are the token contracts. We next further investigate the token contracts. In order to collect the information of tokens, we process the receipt dataset to extract the standard events, which are defined in the standard ERC20 protocol of Ethereum community \cite{buterin2015erc20}. Additionally, each ERC20 token contains basic information like name, symbol, total supply, etc. We then send calls to the local Ethereum peers to collect such basic information of ERC20 tokens.

As shown in Table~\ref{Statistics of Dataset 5}, 106,683 smart contracts are considered as ERC20 contracts, since they output the events that are defined as the standard ERC20 token transactions. There are 227,698,645 ERC20 transactions among 42,146,575 holder addresses. Generally, the number of holder addresses could be much more larger than that of exact human holders because a user may own several addresses. Meanwhile some token issuers will send the tokens to other users without their permissions (also called \emph{token air-drop} \cite{van2017token}). 

\begin{figure}
\centering 
\subfigure[ERC20 Popularity Distribution]{
\centering
\includegraphics[width=4.1cm]{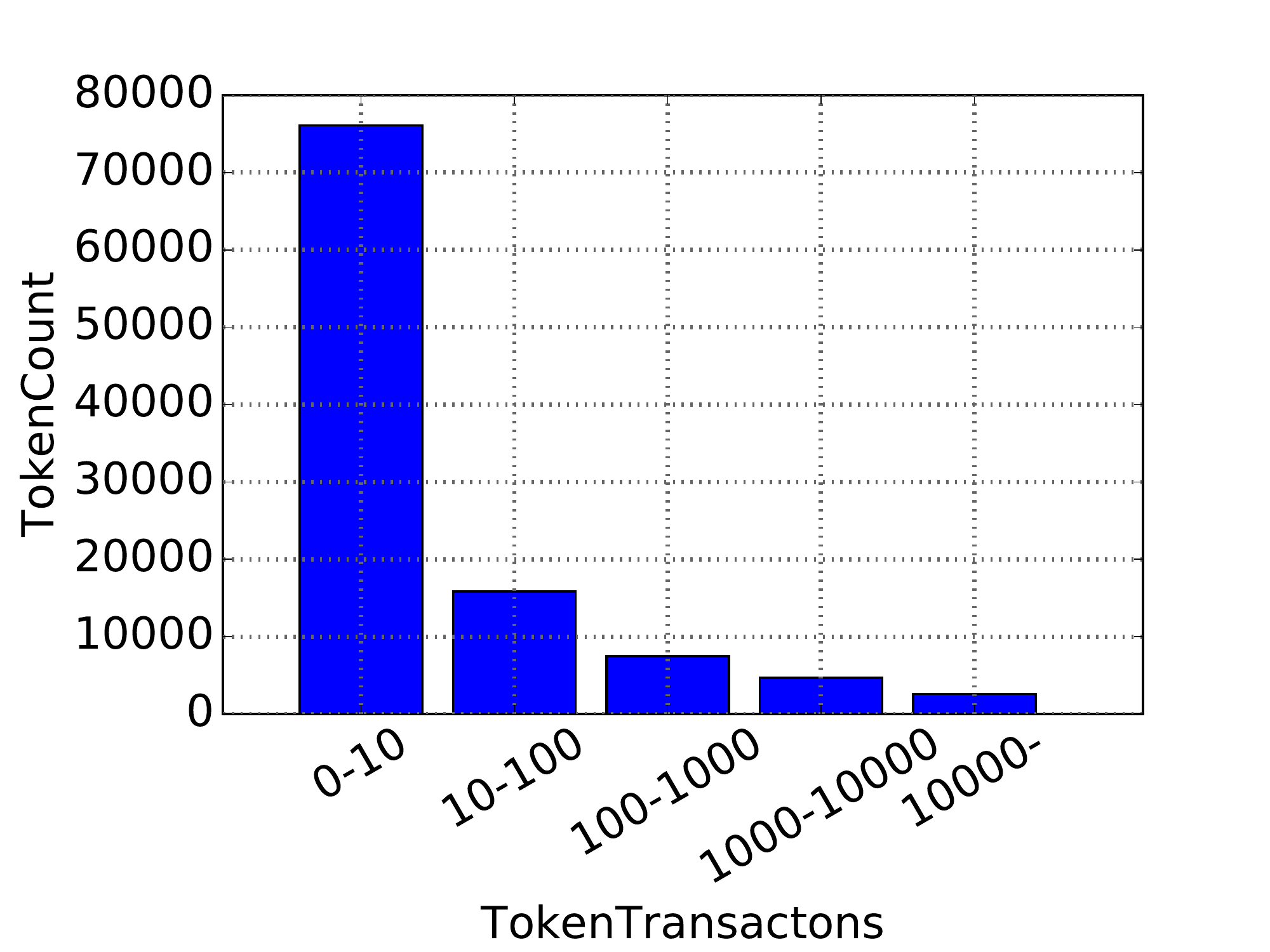} \label{5_erc20count}
}
\subfigure[Word Cloud of ERC20 Tokens]{
\centering
\includegraphics[width=4.1cm]{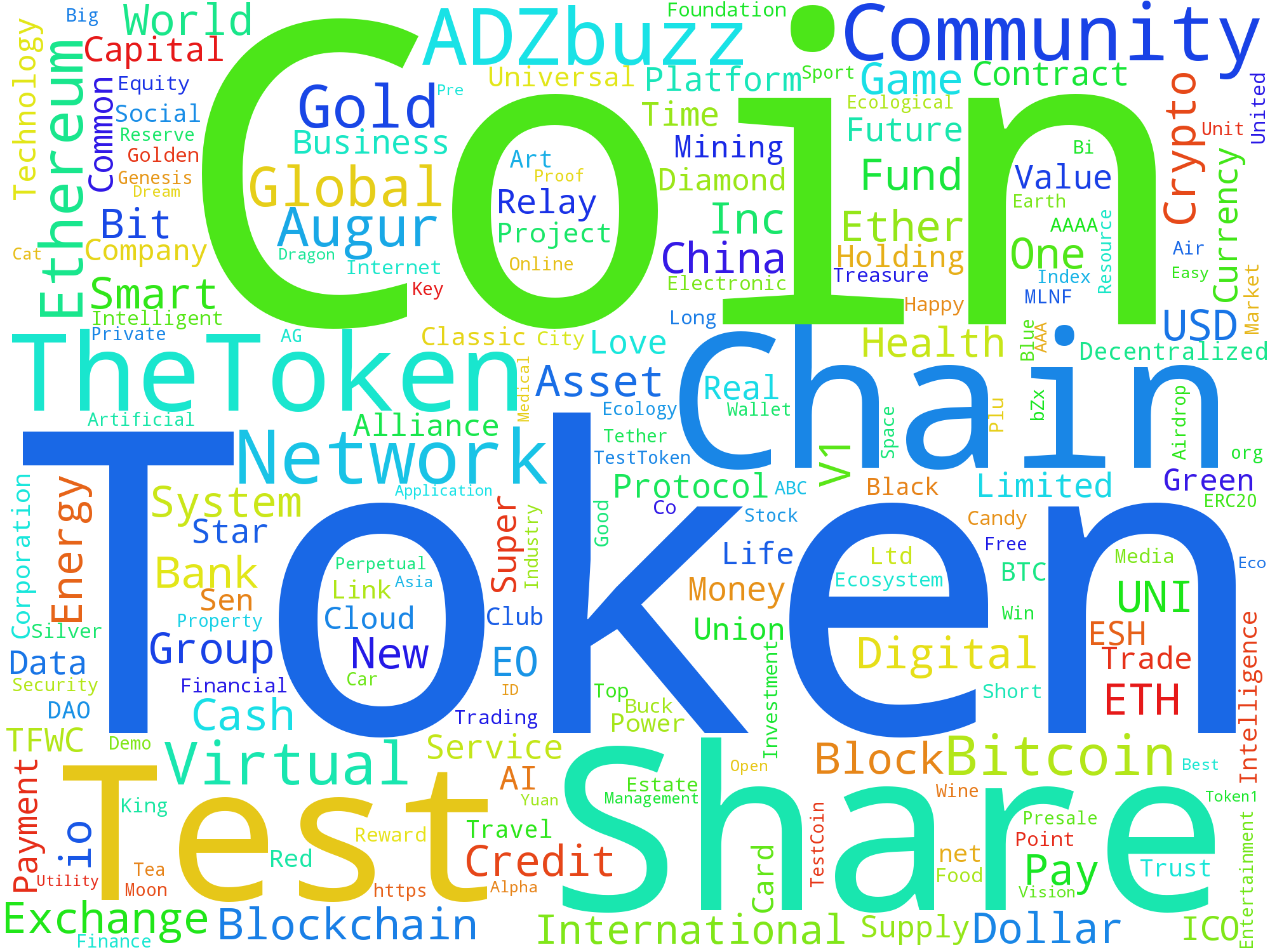} \label{5_erc20Cloud}
} 
\caption{Visualization of Dataset 5}
\end{figure}

Figure~\ref{5_erc20count} shows the transaction count distribution for each ERC20 token. We can easily observe the Matthew effect \cite{merton1968matthew} from Figure~\ref{5_erc20count} as most of token transactions happen in few token contracts. Figure~\ref{5_erc20Cloud} presents the word cloud of names of ERC20 tokens. It is shown in Figure~\ref{5_erc20Cloud} that the most common words are ``Chain'', ``Coin'', and ``Share'', on which the most ERC20 tokens focus. In addition, another common word is ``Test'', implying that many ERC20 contracts deployed on Ethereum are just for the testing purpose.

\begin{table}[]
    \centering
    \caption{Statistics of Dataset 5}
    \begin{tabular}{|l|r|}
    \hline
    \textbf{Statistics} &  \textbf{Values} \\ \hline\hline
    No. of ERC20 Contracts     &      106,683\\ \hline    
    No. of ERC20 Transactions     &  227,698,645   \\ \hline    
    No. of Holder Addresses     &   42,146,575\\ \hline  
    \end{tabular}
    \label{Statistics of Dataset 5}
\end{table}

\renewcommand\subfigcapskip{-0.5ex}

\subsection{Dataset 6: ERC721 Token Trasnaction}\label{Dataset 6: ERC721 Token Trasnaction}

ERC721 token is another contract protocol proposed by Ethereum community \cite{entriken2018erc}. Different from ERC20 token, ERC721 token is indivisible. In the contract function, the parameter is not the value of token but the token ID. For example, a virtual pet in smart contract could be a ERC721 token, which is not separable but can be transferred. 

\begin{figure}
\centering 
\subfigure[ERC721 Popularity Distribution]{
\centering
\includegraphics[width=4.1cm]{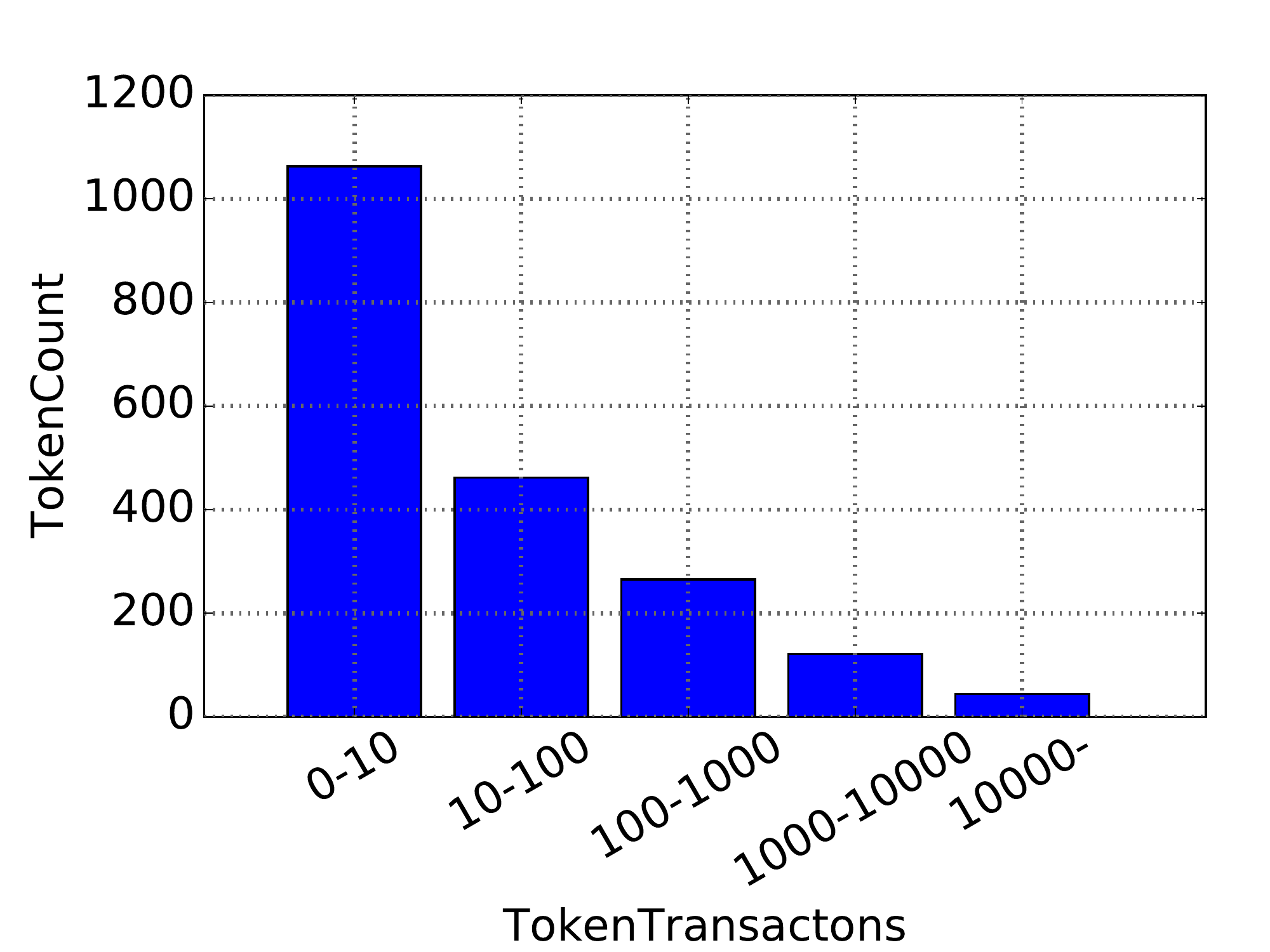} \label{6_erc721count}
}
\subfigure[CryptoKitties Turnover Times]{
\centering
\includegraphics[width=4.1cm]{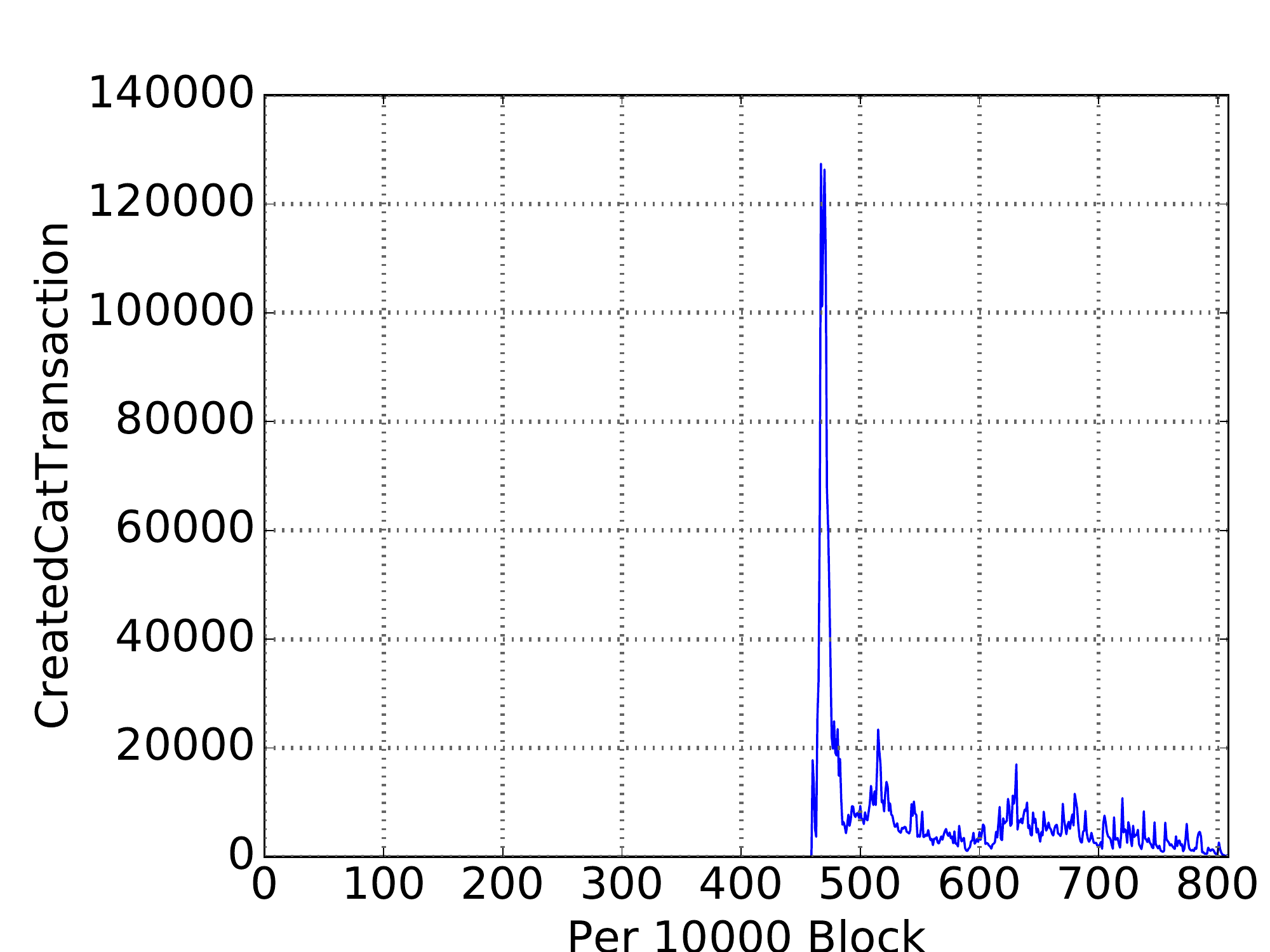} \label{6_catcount2}
} 
\caption{Visualization of Dataset 6} \label{cost}
\end{figure}

Table~\ref{Statistics of Dataset 6} presents the statistics of ERC721 contracts. We find that 1,954 ERC721 contracts contain 7,524,827 token transactions and 414,829 holder addresses. It is worth mentioning that some of the collected contracts do not follow the standard ERC721 protocol exactly. These contracts are also included in the dataset since they output the token transferred events in the receipt. Figure~\ref{6_erc721count} shows the popularity distribution of ERC721 tokens. Compared with ERC20 tokens, the amount of ERC721 tokens is much lower. The major reason is that ERC721 applications require much more workloads on visualization at each token, consequently improving the development difficulty.

We also investigate a popular ERC721 token contract called CryptoKitties. It is one of the most famous ERC721 token contracts, selling the virtual cats as tokens. Each cat is represented as a token in the ERC721 contract. We count the turnover times distributed by birth block of the cats, as shown in Figure~\ref{6_catcount2}. Figure~\ref{6_catcount2} also shows that the cats that were born in 4,500,000 to 5,000,000 blocks have the higher turnover times than others. At that time, the type of CryptoKitties reaches the peak. The time to obtain the peak in Figure~\ref{6_catcount2} is almost the same as that in Figure~\ref{1_txcount} and Figure~\ref{1_gasprice}, implying that the popularity of CryptoKitties leads to the congestion of Ethereum.

\begin{table}[]
    \centering
    \caption{Statistics of Dataset 6}
    \begin{tabular}{|l|r|}
    \hline
    \textbf{Statistics} &  \textbf{Values} \\ \hline\hline
    No. of ERC721 Contracts     &  1,954   \\ \hline    
    No. of ERC721 Transactions     & 7,524,827   \\ \hline    
    No. of Holder Addresses     &  414,829 \\ \hline  
    \end{tabular}
    \label{Statistics of Dataset 6}
\end{table}

\renewcommand\subfigcapskip{-0.5ex}

\section{Applications of XBlock-ETH}
\label{sec:application}

This section presents applications of XBlock-ETH framework. As shown in Figure~\ref{0_arc}, the architecture of Ethereum consists of peers, blockchain, smart contracts and tokens. Thus, we also categorize the applications according to top-3 layers (i.e., blockchains, smart contracts and tokens). Meanwhile, we also discuss the research opportunities in each layer.

\subsection{Blockchain System Analysis}
Since XBlock-ETH processes data from realistic blockchain systems, it can be used to support the following applications.

\subsubsection{Decentralization Analysis}
The decentralization is one of the key features of blockchain systems. However, there are few studies on the decentralization evaluation of the blockchain systems. In particular, the work of \cite{wang2019measurement} presents the measurement of the mining pool for Bitcoin. Although Gencer et al. \cite{gencer2018decentralization} present a measurement study on the decentralization level of Bitcoin and Ethereum, their study only consider several metrics such as network bandwith, mining power and fairness. In contrast, our XBlock-ETH data offers a more comprehensive measurement on Ethereum. Moreover, our work can be used to analyze the decentralization of users, contract owners and miners. In addition, our XBlock-ETH can also be used to make comparison with other blockchain systems, such as Bitcoin, EOS or other blockchain systems.

\subsubsection{Gasprice Prediction}
Since the transaction fees are equal to ``\texttt{gasPrice}'' times ``\texttt{gasUsed}'', the users can control the ``\texttt{gasUsed}'' in a reasonably low range to minimize the transaction fees charged by miners. Meanwhile, we can learn from Section~\ref{Dataset 1: Block and Transaction} that there is always a gap between the minimum ``\texttt{gasPrice}'' and the average ``\texttt{gasPrice}'' in a block, leading to the opportunity to save fees. Recent studies such as Other-tech\cite{gas1}, Gitcoin\cite{gas2}, Majuri\cite{gas3} analyze the ``\texttt{gasPrice}'' of Ethereum while several Ethereum websites (e.g., Etherscan\footnote{http://etherscan.io\label{http://etherscan.io}}, Etherchain\footnote{http://etherchain.org}) provide tools to predict the ``\texttt{gasPrice}'' in a short time. However, those tools are essentially \emph{black boxes} and the accuracy of them cannot be assured. In summary, the prediction of ``\texttt{gasPrice}'' has great economic value such that the user of Ethereum can save the money or shorten waiting time through the ``\texttt{gasPrice}'' prediction while it is worthwhile to conduct an in-depth study in the future.

\subsubsection{Performance Benchmark}
Performance is crucial to blockchain systems. There are a number of studies on blockchain performance optimizations, such as Omniledger\cite{kokoris2018omniledger}, Algorand\cite{gilad2017algorand} and RapidChain\cite{zamani2018rapidchain}. Meanwhile, some optimized blockchain systems (e.g., Monoxide\cite{wang2019monoxide}) adopt the realistic blockchain transaction data to conduct performance evaluation for blockchain systems. To compare the performance of different optimization methods, a common benchmark of real-world user cases for blockchain systems is needed. Zheng et al.\cite{zheng2018detailed} and BlockBench \cite{dinh2017blockbench} propose the performance evaluation of blockchain systems. The performance benchmark requires simulating the user behaviors and obtaining data similar to real-world blockchain systems. In this aspect, the XBlock-ETH framework can be regarded as a benchmark since the source data is generated exactly by the real-world users.

\subsection{Smart Contract Analysis}
As one of the most popular smart contract platforms, Ethereum has attracted a large number of software developers as well a huge number of smart contracts. Therefore, Ethereum has a more active developer community compared with other smart contract platforms such as EOS and Tron, which claim to have the higher throughput and lower latency than Ethereum. Consequently, our XBlock-ETH framework (on top of Ethereum) can be used in the studies of smart contracts. We summarize the potential applications of XBlock-ETH as follows.

\subsubsection{Contract Similarity and Recommendation}
As indicated in Section~\ref{sec:dataexploration}, there is a great similarity between the smart contract codes and call of smart contracts. Code similarity evaluation is a traditional research topic in software engineering as a number of studies concentrate on code similarity detection \cite{chilowicz2009syntax}\cite{luo2014semantics}\cite{lemhofer2004recognizing}. Several recent studies focus on similarity analysis of smart contracts. In particular, Etherscan\textsuperscript{\ref{http://etherscan.io}} provides the query system based on similar contracts. Finding the similar contracts is beneficial to the developers during developing new contracts. For example, developers can estimate the user behaviors before the publishing the contract. Meanwhile, Huang et al.\cite{huang2019recommending} propose the method to recommend differentiated codes to update smart contracts based on the existing codes of smart contracts. In addition, in the aspect of users, recommending the similar smart contract will help users to find the contracts suitable for themselves.

\subsubsection{Contract Developer Analysis}
Developer analysis that is another traditional research topic in software engineering includes developer network analysis\cite{meneely2008predicting}, behavior analysis \cite{layman2007toward}, fault prediction \cite{weyuker2007using}, and so on. With respect to developer analysis, XBlock-ETH also includes a large network of smart contract developers. For example, there some on-chain libraries deployed and provided by different developers; these libaries  can be invoked by others. Each developer can be identified by his/her own Ethereum address. Thus, the contract calling network can be also regarded as the collaboration network of contract developers. The network and structure of developer collaboration may inform us about the reliability of the contract codes. For example, the developer who develops a smart contract with vulnerabilities will have a higher risk to develop new contracts with vulnerabilities than others. In this sense, our XBlock-ETH can be beneficial to the developer analysis after analyzing smart contracts of developers.

\subsubsection{Contract Vulnerability Detection} \label{Contract Vulnerability Detection}
The security of smart contracts has been a hot research topic in blockchain research community. In particular, the vulnerability of smart contracts has attracted extra attentions. A number of malicious attacks on Ethereum (e.g., TheDAO attack) have already resulted in huge loss (in terms of tens of millions of dollars) \cite{mehar2019understanding}. To prevent smart contracts from malicious attacks, the vulnerability detection on contracts is a critical step. There are some recent attempts in vulnerability detection. For example, Oyente \cite{luu2016making}, Zeus \cite{kalra2018zeus}, teEther \cite{krupp2018teether}, S-gram \cite{liu2018s}, ContractFuzzer \cite{jiang2018contractfuzzer} propose the tools of vulnerability detection on smart contracts. In some cases, the vulnerability detection methods of smart contracts can be inspired and motivated by traditional software vulnerability detection methods as they are essentially equivalent to the verification of the codes. In this aspect, several studies focus on verifying contract codes on blockchains; these contract codes are also called ``bytecode'' or ``opcode''. Our XBlock-ETH that essentially includes the data of contract codes can be applied to contract vulnerability detection.

\subsubsection{Fraud Detection}
Due to the huge economic value and the popularity of smart contracts, smart contracts can be exploited by malicious users as scams. For example, crowd-funding contracts with a promised huge return to attract victims for investment. It is reported in \cite{Weili18} that Ponzi scam contracts can defraud others' cryptocurrencies. Several approaches \cite{Weili18,bartoletti2019dissecting,chen2019exploiting,Torres:2019} have been proposed to detect the fraud contracts on Ethereum. Most of the methods are mainly based on the codes and transaction records of smart contracts while they are included in XBlock-ETH data. Thus, XBlock-ETH data can be further leveraged in fraud detection.

\subsection{Cryptocurrency Analysis}
Blockchain-based cryptocurrency has become a hot topic recent years due to the decentralization and the reduced cost. There are a large amount of cryptocurrencies in Ethereum, including the Ether, ERC20 tokens and ERC721 tokens. It is shown in the CoinMarketCap\footnote{https://coinmarketcap.com/all/views/all/} that more than 2,000 kinds of tokens can be used in third-party exchange. Therefore, cryptocurrency analysis based on blockchain data can bring huge financial values. We roughly categorize the cryptocurrency analysis into cryptocurrency transferring analysis, cryptocurrency price analysis and fake user detection, which are explained as follows.

\subsubsection{Cryptocurrency Transferring Analysis}
Analysis on cryptocurrency transactions is a preliminary step to conduct cryptocurrency transferring analysis. Regarding Ether transferring, Chen et al. \cite{chen2018understanding} propose the graph analysis on Ether transactions and derive some insights from graph analysis. With regard to ERC20/ERC721 tokens, Victor et al.\cite{victor2019measuring} and Somin et al.\cite{somin2018network} propose the analysis of the token trading network. After the analysis on cryptocurrency transactions, the further analysis on user behaviours can be done. For example, the users of tokens may form different communities. The community discovery can be conducted through analyzing cryptocurrency transactions. Moreover, the anonymity of blockchain-based cryptocurrency can result in money-laundering behaviors, which can be essentially identified and detected via cryptocurrency transaction analysis. Our XBlock-ETH data offers the potential solutions to these issues.

\subsubsection{Cryptocurrency Price Analysis}
The price of blockchain-based cryptocurrencies has been affected by multiple different factors such as government policies, technology innovations, social sentiment and business activities. Several recent studies focus on the price analysis and prediction of cryptocurrencies \cite{lamon2017cryptocurrency, abraham2018cryptocurrency, mensi2019structural}. The typical cryptocurrency price analysis consists of three steps: (i) collect price data form the cryptocurrency exchanges, (ii) identify the relevance between cryptocurrency prices and other factors, (iii) forecast the future prices and predict the potential profits. However, the price of cryptocurrencies can sometimes be maliciously controlled by some parties. Thus, the data cleaning process is necessary to obtain the accurate and normal cryptocurrency price data. Our XBlock-ETH also contains cryptocurrency price data, which can be used for cryptocurrency price analysis while the raw receipt data may require the further preprocess to benfit the future analysis.


\subsubsection{Fake User Detection}
Fake user detection \cite{cao2012aiding, varol2017online, ferrara2016rise} is a traditional research topic in social networks. The cryptocurrency users in blockchain systems also form social-network like communities, in which there are also some fake users controlled by the developers to improve the DApps activity rankings. Because the DApp (or cryptocurrency) ranking is based on some metrics related to the user activities, such as Daily Active Users (DAU). Therefore, many developers exploit the loophole to fabricate some fake users to improve activities so as to gain higher rankings. Although some DApp websites, such as DAppReview\footnote{http://dapp.review } mark the cryptocurrencies with fake users, this kind of fake user detection is almost done in a black box or manually. In addition, there are few studies on fake user detection on cryptocurrency. The permission-less blockchain systems which are often free of charge may advocate more frequent fake user activities than permissioned  blockchain systems. Our XBlock-ETH will be further improved to support the fake user detection in the future.

\section{Related Work and Discussion} 
\label{sec:related}
Some previous studies on Ethereum data will be described and discussed in this section. We categorize the state-of-the-art literature into two types: \textit{Data tools} and \textit{Data analysis}.

Regarding Ethereum data tools, some studies provide open-source tools or APIs with users to obtain the data. For example, EtherQL \cite{li2017etherql} offers a query layer for Ethereum. Blocksci \cite{kalodner2017blocksci} constructs a platform for researchers to analyze the blockchain data. DataEther \cite{chen2019dataether} is a tool to obtain the data from Ethereum, with code modification of the Ethereum clients. Google BigQuery \cite{tigani2014google} imports the data of Bitcoin and Ethereum and enables researchers to analyaze the data online while updating Ethereum data has been stopped for a long time. Meanwhile, it is pretty challenging for researchers to download, update and analyze the blockchain data. There are also some websites offering data APIs for developers to use or analyze, including Amberdata\footnote{http://amberdata.io}. However, these third-party APIs always restrict the usage rating so that it is difficult for researchers to crawl all the data. In summary, most of these studies only offer tools or APIs to researchers while failing to offer well-processed up-to-date datasets.

Some recent studies provide the analysis on the Ethereum data. For example, studies of \cite{Weili18,chen2019exploiting,bartoletti2019dissecting} propose the contract classification  methods to detect Ponzi schemes. Moreover, Chen et al. \cite{chen2018understanding} analyze the transactions and construct three graphs to observer the behaviors on Ethereum. Furthermore, the work of \cite{tokenscope} analyzes the ERC20 tokens on Etherem and find un-standard token. Another popular research area on Ethereum data is the smart contracts security. For example, Oyente \cite{luu2016making}, Zeus\cite{kalra2018zeus} propose the security analysis tools for Ethereum smart contracts to find the vulnerable codes. Although some of these studies release some datasets, most of them are only suitable for specific research questions. Furthermore, most of them are difficult to be updated.

It is worth mentioning that XBlock-ETH does not contain the off-chain data such as the price data in exchanges, the source code of verified smart contracts, the behavior on Github of the DApps even if they are also crucial for the analysis. Since those data are not generated by the Ethereum, we only concentrate on the on-chain data in this paper. 

\section{Conclusion and Future work} \label{sec:conc}

This paper introduces a well-processed up-to-date on-chain dataset of Ethereum, namely XBlock-ETH, which includes the data of the Ethereum blockchain, smart contracts and cryptocurrencies. Moreover, comprehensive statistics and exploration of the datasets are presented. The XBlock-ETH datasets have been released on XBlock.pro website. Furthermore, the research opportunities of the XBlock-ETH datasets are also outlined. 

Our XBlock-ETH is promising to promote the studies on Ethereum. The future improvements are listed as following: 
\textbf{(1) More features:} The exploration of the basic features of the datasets are given in this paper. Ethereum is a complex ecosystem that includes decentralized finance, stable coin, and so on. More features of the Ethereum data will be explored in the future. 
\textbf{(2) More data from exchanges and open-source communities:} The off-chain data is also important since it provides the information of off-chain behaviors of both developers and users. In the future, the off-chain data will be collected. 
\textbf{(3) Combined analysis with other blockchain systems:} There are some other blockchain systems that have also attracted a large number of users and developers. The combined analysis between Ethereum and other permission-less blockchains will be conducted in the future.

\section*{Acknowledgment}
The work described in this paper was supported by TBD.

\newpage
\bibliographystyle{IEEEtran}
\bibliography{IEEEabrv,yinyong}

\begin{IEEEbiography}[{\includegraphics[width=1in]{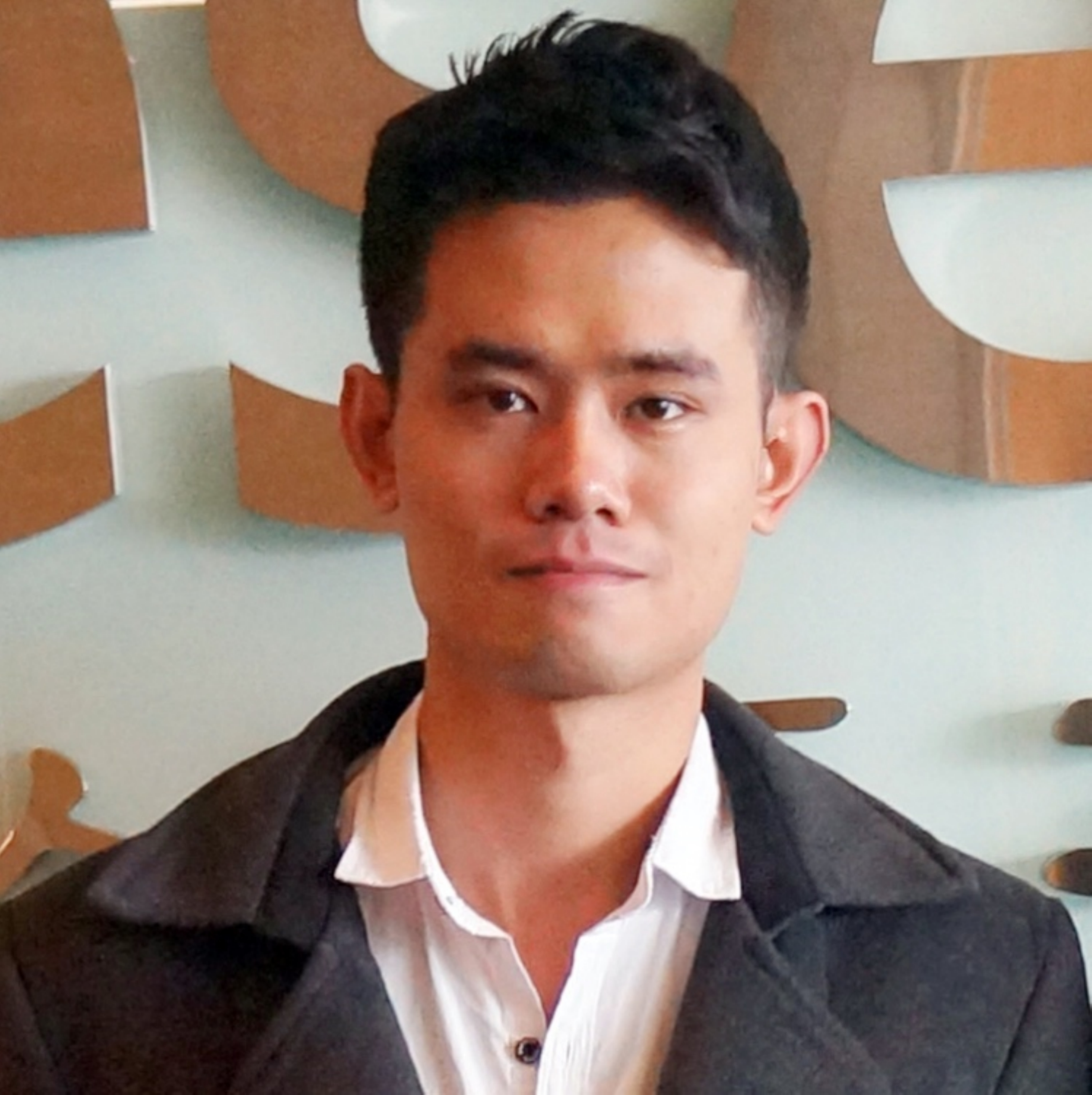}}]{Peilin Zheng} is a student at Sun Yat-sen University, Guangzhou, China. His research interests include performance monitoring and evaluation on blockchain, optimization of smart contracts, and blockchain-based decentralized applications.
\end{IEEEbiography}

\begin{IEEEbiography}[{\includegraphics[width=1in]{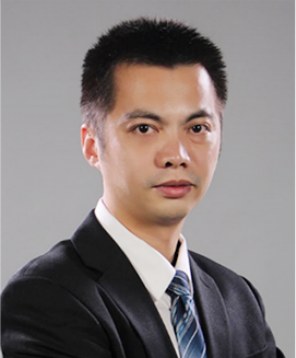}}]{Zibin Zheng} is a professor at Sun Yat-sen University, Guangzhou, China. He received Ph.D. degree from The Chinese University of Hong Kong in 2011. He received ACM SIGSOFT Distinguished Paper Award at ICSE' 10, Best Student Paper Award at ICWS' 10, and IBM Ph.D. Fellowship Award. His research interests include services computing, software engineering, and blockchain.
\end{IEEEbiography}

\begin{IEEEbiography}[{\includegraphics[width=1in]{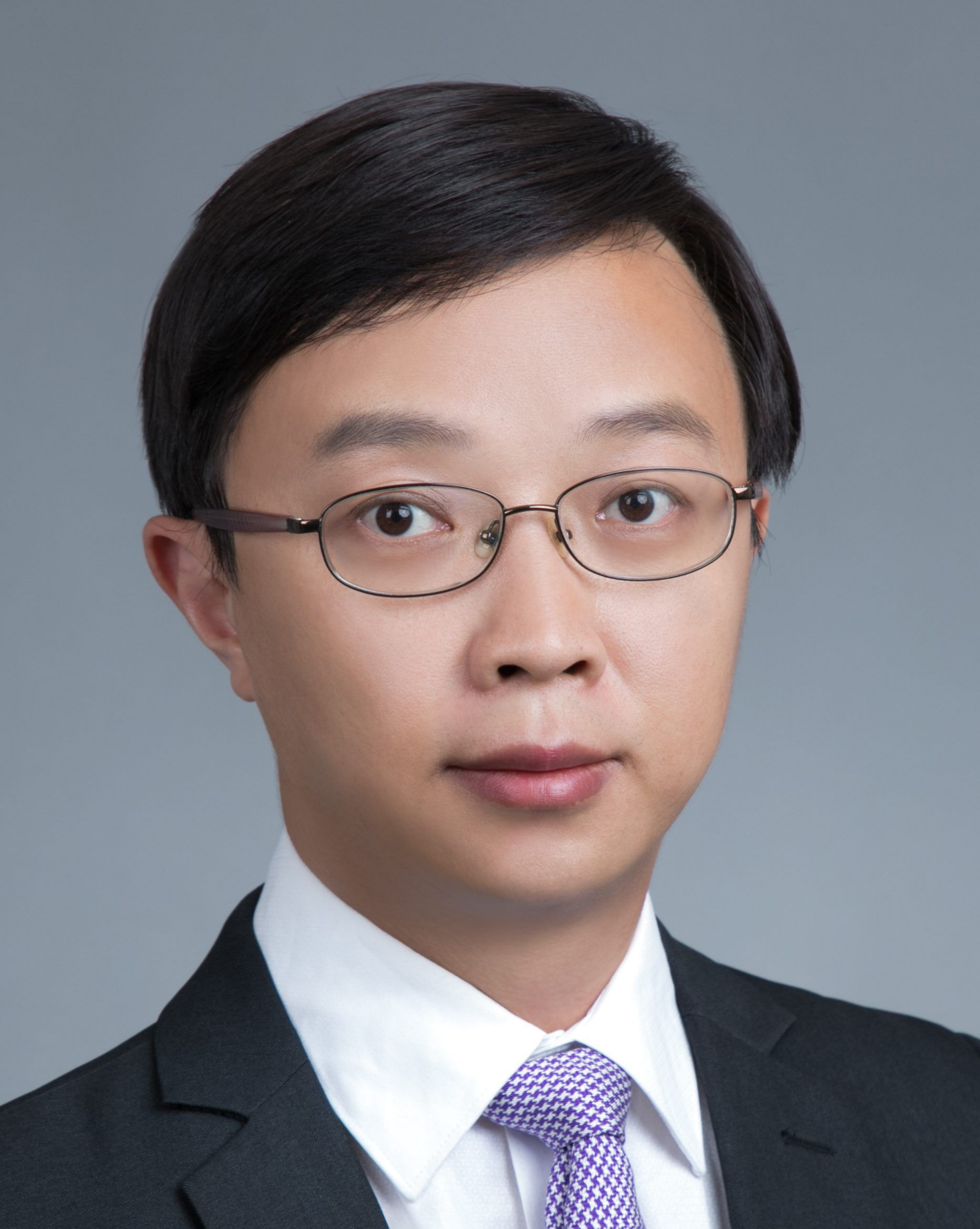}}]{Hong-Ning Dai} is an Associate Professor in Faculty of Information Technology at Macau University of Science and Technology. He obtained his PhD in Computer Science and  ngineering from the Department of Computer Science and Engineering at the Chinese University of Hong Kong in 2008. His research interests include wireless networks, mobile computing, and distributed systems.
\end{IEEEbiography}

\vspace{12pt}
\end{document}